%% file: 3PCF_VIPERS_final.tex
\documentclass[longauth]{aa}

\usepackage{amsmath,amssymb}

\usepackage{graphicx}
\usepackage{txfonts}
\usepackage{subfigure}
\usepackage{float}
\usepackage{natbib,twoopt}
\usepackage{xspace}
\usepackage{balance}
\usepackage{upgreek}
\usepackage{multirow}
\usepackage[colorlinks=true,citecolor = blue,]{hyperref}

\def\apjl{Astrophys. J. Lett.}
\def\apjs{Astrophys. J. Suppl.}

\def\aap{Astron. Astrophys.}

\newcommand{\Mpch}{$h^{-1}\,\mbox{Mpc}$\,}

\bibpunct{(}{)}{;}{a}{}{,} 
\DeclareGraphicsExtensions{.eps,.ps,.pdf}

\begin{document}

\title{The VIMOS Public Extragalactic Redshift Survey (VIPERS)
  \thanks{Based on observations collected at the European Southern
    Observatory, Paranal, Chile under programmes 182.A-0886 (LP) at the
    Very Large Telescope, and also based on observations obtained with
    MegaPrime/MegaCam, a joint project of CFHT and CEA/DAPNIA, at the
    Canada-France-Hawaii Telescope (CFHT), which is operated by the
    National Research Council (NRC) of Canada, the Institut National
    des Science de l'Univers of the Centre National de la Recherche
    Scientifique (CNRS) of France, and the University of Hawaii. This
    work is based in part on data products produced at TERAPIX and the
    Canadian Astronomy Data Centre as part of the Canada-France-Hawaii
    Telescope Legacy Survey, a collaborative project of NRC and
    CNRS. The VIPERS web site is http://vipers.inaf.it/.}  }

\subtitle{Exploring the dependence of the three-point correlation
  function on stellar mass and luminosity at $0.5<z<1.1$}

\include{authors}

\date{Received --; accepted --}

\abstract 
{}
{The three-point correlation function (3PCF) is a powerful probe to
  investigate the clustering of matter in the Universe in a
  complementary way with respect to lower-order
  statistics, providing additional information with respect to the two-point correlation function  
  and allowing us to shed light on biasing, nonlinear processes, and deviations from Gaussian statistics. 
  In this paper, we analyse the first data release of the
  VIMOS Public Extragalactic Redshift Survey (VIPERS), determining 
  the dependence of the three-point
  correlation function on luminosity and stellar mass at $z=[0.5,1.1]$.}
{We exploit the VIPERS Public Data Release 1, consisting of more than
  50,000 galaxies with B-band magnitudes in the range $-21.6\lesssim M_{\rm B}-5\log(h)\lesssim-19.9$
  and stellar masses in the range $9.8\lesssim
  \log(M_\star[h^{-2}\,M_\odot])\lesssim 10.7$. We measure both the
  connected 3PCF and the reduced 3PCF
  in redshift space, probing different configurations and scales, in
  the range $2.5<r\,$[\Mpch]$<20$.}
{We find a significant dependence of the reduced 3PCF on scales and
  triangle shapes, with stronger anisotropy at larger scales ($r\sim10$ \Mpch) and an almost flat trend at smaller scales, $r\sim2.5$ \Mpch.
  Massive and luminous galaxies present a larger connected 3PCF,
  while the reduced 3PCF is remarkably insensitive to magnitude and stellar masses in the range we explored. 
  These trends, already observed at low redshifts, are confirmed for the first time to be still valid up to $z=1.1$,
  providing support to the hierarchical scenario for which massive and bright systems are expected to be more clustered. 
  The possibility of using the measured 3PCF to provide independent constraints on the linear galaxy bias $b$ has also been explored,
  showing promising results in agreement with other probes.
  }
{}

    
\authorrunning{M.Moresco et al.}  \titlerunning{Dependence of the
  three-point correlation function on stellar mass and luminosity}

\maketitle

\section{Introduction} 
The clustering of galaxies, and its evolution with cosmic time, is one
of the major probes of modern cosmology. It provides crucial
constraints on the underlying distribution of matter in the Universe,
of which galaxies represent a biased tracer, and thus helps improve
our knowledge of the fundamental components driving the evolution of
the Universe. In particular, the two-point correlation function (2PCF)
and the power spectrum $P(k)$, its analogous in Fourier space, have
been extensively exploited as cosmological probes.  Encoded in the
shape and amplitude of the two-point statistics is the imprint of
primordial fluctuations and their evolution in the pre- and
post-recombination era, the most notable example of which is the
baryonic acoustic oscillation (BAO) signature that, for the 2PCF, is
in the form of a peak at a characteristic length scale that can
provide an ideal ``standard ruler''.  Indeed, this feature has been
exploited to place constraints on the expansion history of the
Universe, measuring both the Hubble parameter $H(z)$ and the angular
diameter distance $D_{A}(z)$, offering new insights on the nature of
dark energy and dark matter \citep{eisenstein2005, cole2005,
  blake2011,beutler2011, anderson2012, anderson2014, cuesta2016}.

Galaxy clustering can provide useful information also to
understand how galaxies have evolved with cosmic time. In particular,
it has been found that more luminous and massive galaxies are more
strongly clustered than fainter and less massive ones
\citep{davis1976, davis1988, hamilton1988, loveday1995, benoist1996,
  guzzo1997, guzzo2000, norberg2001, norberg2002, zehavi2002,
  zehavi2005, zehavi2011, brown2003, abbas2006, li2006, swanson2008,
  ross2011, guo2013, marulli2013}; similar trends have been found
also as a function of morphology and colours, for which galaxies with a rounder shape and redder
show an enhanced clustering. The 2PCF is also often
used to provide constraints on galaxy bias $b$, which quantifies the
excess in clustering of the selected sample with respect to the
underlying dark matter; however, these estimates have to assume a
fiducial cosmology, and are degenerate with the amplitude of linear
matter density fluctuations quantified at 8 \Mpch, $\sigma_{8}$, of the assumed
model.

While a Gaussian field can be completely described by its two-point
statistics, to detect non-Gaussian signals, both of primordial type and induced by non-linear
evolution of clustering, and to understand the
evolution of matter beyond the linear approximation, it is necessary
to study higher-order statistics. The first significant
order above 2PCF and power spectrum is represented by the three-point
correlation function (3PCF) and bispectrum $B(k)$ respectively.  These functions
provide complementary information with respect to lower-order
statistics, and can be used in combination with them to break
degeneracies between estimated cosmological parameters, such as galaxy
bias and $\sigma_{8}$.  Many studies have exploited
both the evolutionary and cosmological information encoded in the
three-point statistics, both in configuration space
\citep[e.g.][]{fry1994, frieman1994, jing1997, jing2004, kayo2004,
  gaztanaga2005, nichol2006, ross2006, kulkarni2007, mcbride2011b,
  mcbride2011, marin2011, marin2013, guo2014, moresco2014, guo2015}
and in Fourier space \citep[e.g.][]{fry1982, matarrese1997, verde1998,
  verde2000, scoccimarro2000, scoccimarro2001, sefusatti2005,
  sefusatti2007}.

The aim of this paper is to push these investigations to higher redshifts.
For this purpose, we analyse the VIPERS Public Data Release 1
(PDR-1) \citep{guzzo2014, garilli2014}, constraining the dependence of
the 3PCF on stellar mass and luminosity. A similar analysis of the
same dataset has been performed by \cite{marulli2013} (hereafter M13),
but for the 2PCF.  This paper is intended as an extension of the
analysis done in M13, exploiting the additional constraints that can
be obtained from higher-order correlation functions.
In particular, we focus our analysis on the non-linear or mildly non-linear evolution 
regime, since the size of the survey does not allow us to probe those scales 
sensitive to possible primordial non-Gaussianities. 

The paper is organized as follows. In Sect. \ref{sec:data} we
present the VIPERS galaxy sample, and describe how it has been divided
into stellar mass, luminosity, and redshift sub-samples. In Sect.
\ref{sec:method} we discuss how the 3PCF and its errors have been
measured. Finally, in Sect. \ref{sec:results} we present our
results, showing the dependence of the measured 3PCF on scales,
triplet shapes, redshifts, luminosity, and the stellar mass, and the estimate of galaxy bias. 
In Appendix \ref{appendix}, we provide information on the covariance matrices for 
our analysis

Throughout this paper, we adopt a standard flat $\Lambda$ cold dark matter (CDM) 
cosmology, with $\Omega_{M}=0.25$ and $h=H_{0}/100=0.73$ km s$^{-1}$
Mpc$^{-1}$. Magnitudes are given in the AB system.

\begin{figure*}
  \includegraphics[width=0.49\textwidth]{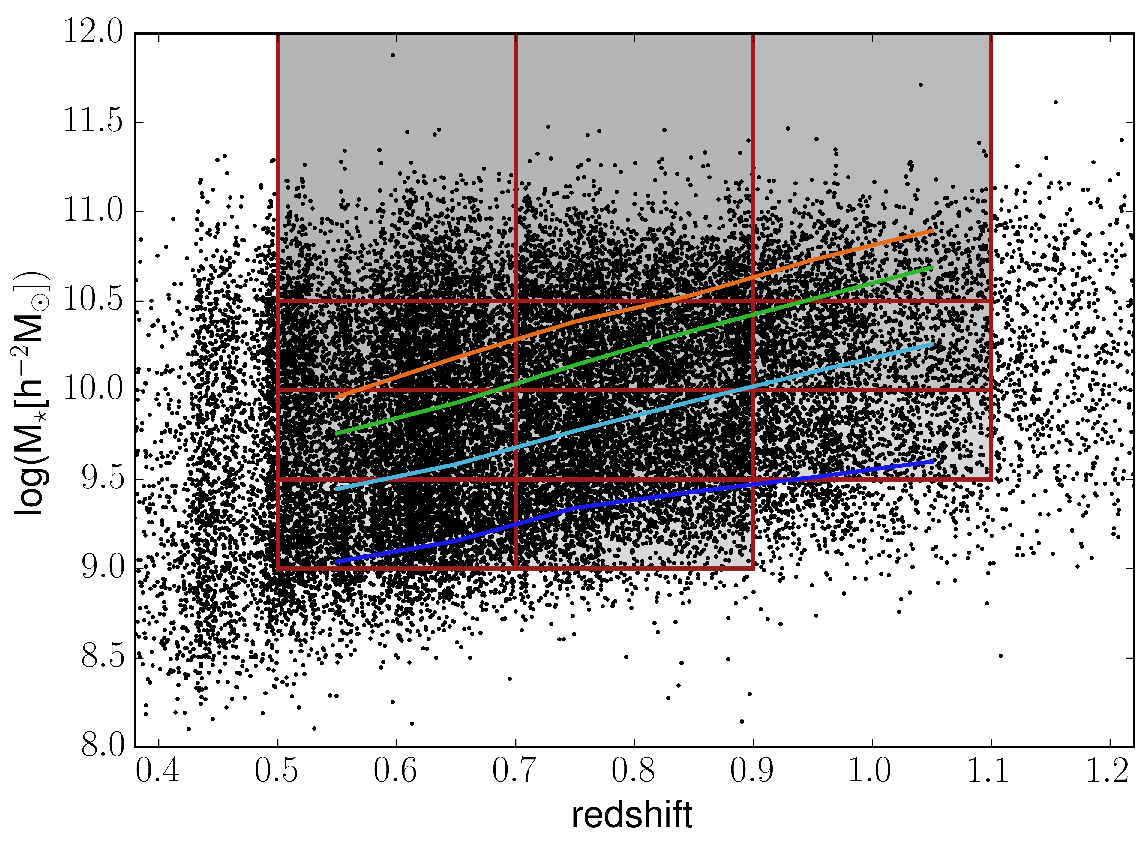}
  \includegraphics[width=0.49\textwidth]{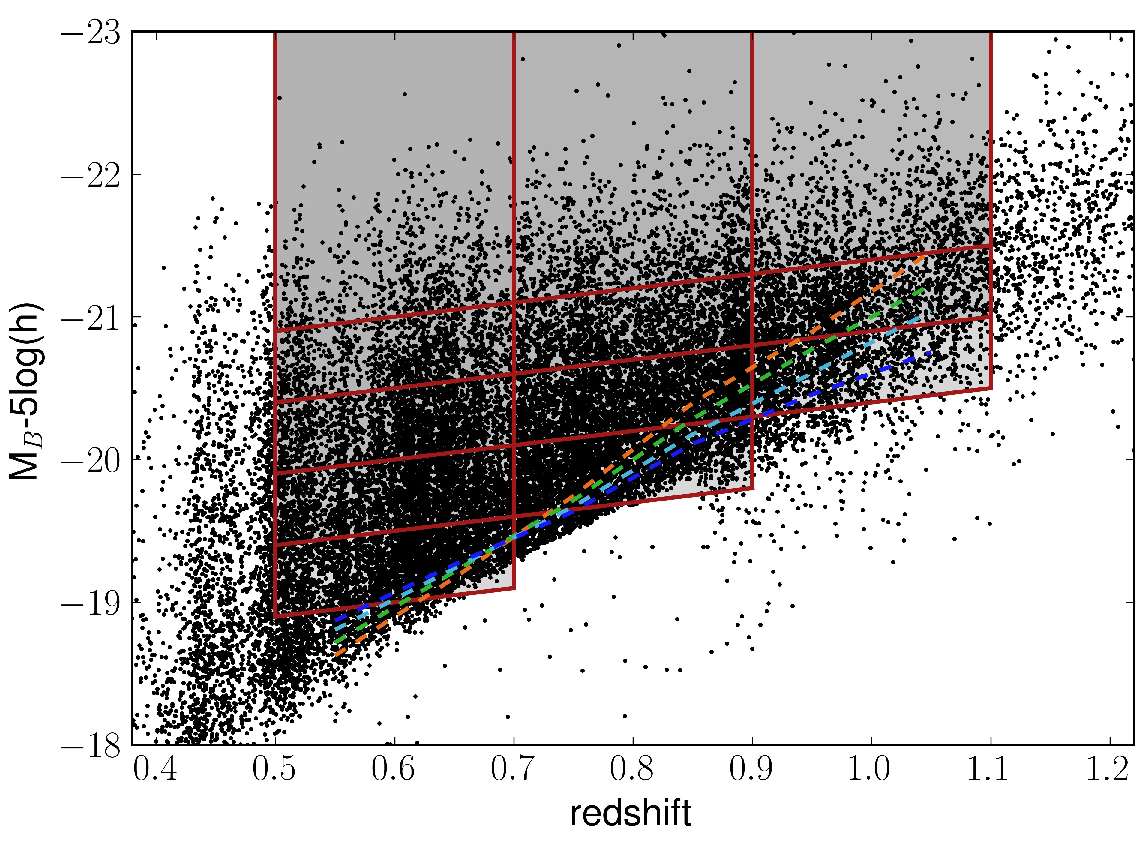}
  \caption{Luminosity- and stellar mass-redshift relations.  The black
    points represent VIPERS galaxies, where red boxes show the
    selected subsamples. The boxes in $M_{\rm B}$ absolute magnitude
    are constructed to follow the redshift evolution of galaxies, as
    discussed in the text. Coloured lines represent the $90\%$
    completeness limits for different sub-populations: orange lines for
    early-type galaxies, green for early-spirals, cyan for late
    spirals, and blue for irregulars and starbursts, respectively.}
  \label{fig:sub_samples}
\end{figure*}


\section{The data}
\label{sec:data}

VIPERS is a recently completed European Southern Observatory (ESO) Large Programme, which has measured
spectroscopic redshifts for a complete sample with $0.5<z\lesssim1.2$.
Its general aim has been to build a sample of the general galaxy
population with a combination of volume
[$\sim5\times10^{7}($\Mpch$)^{3}$] and spatial sampling
[$10^{-2}-10^{-3}$ $\rm h^{3} Mpc^{-3}$] comparable to state-of-the
art local surveys.  Its science drivers are to provide at these
redshifts reliable statistical measurements of ensemble properties of
large-scale structure (such as the power spectrum and redshift-space
distortions), of the galaxy population (such as the stellar mass function),
and their combination (such as galaxy bias and the role of the
environment).  The VIPERS PDR-1 is extensively described in
\citet{guzzo2014} and \citet{garilli2014}. It has publicly released
the measurements of the first 57,204 objects of the survey,
  comprising 2448 stars, and 54,756 galaxies and active galactic nuclei redshifts.  The
survey targets have been selected from the Canada-France-Hawaii
Telescope Legacy Survey Wide (CFHTLS-Wide) optical photometric
catalogues \citep{mellier2008} over two fields, W1 and W4, 
  covering 15.7 and 7.9 deg$^2$, respectively, for a total area of
  $\sim24$ deg$^2$. The sample has been selected with a magnitude
limit of $i_{AB}<22.5$, and a $(u-g)$-$(r-i)$ colour cut to properly
select the desired redshift range $z\ge0.5$. VIPERS spectra have been
measured with the low-resolution grism mounted on the Visible Multi-Object Spectrograph (VIMOS) at the ESO Very Large Telescope (VLT) \citep{lefevre2002,
  lefevre2003}, providing a moderate spectral resolution ($R=230$) and
a wavelength range of 5500-9500 $\rm{\AA}$. Recently, 
the final VIPERS data release has also appeared \citep{scodeggio2016},
and this will be used in a following analysis to further explore the higher-order 
correlations with better accuracy.

Stellar masses and absolute magnitudes have been computed for the
entire VIPERS sample with the public code {\small HYPERZMASS}
\citep{bolzonella2000, bolzonella2010}, which performs a fit to the
spectral energy distribution (SED) of the galaxies. The B Buser filter
has been considered to calculate the absolute magnitudes \citep{fritz2014}. We refer to
\citet{davidzon2013} for a more extended discussion about the
SED-fitting technique.
Among several other investigations, the VIPERS data have been used to
estimate the relation between baryons and dark matter through the
galaxy bias \citep[M13,][]{ cucciati2014, cappi2015, granett2015,diporto2016}.

Following the approach of M13, we divide our sample into three equally
spaced redshift ranges, $z\in[0.5,0.7]$, $z\in[0.7,0.9],$ and
$z\in[0.9,1.1]$. Each of them is further divided into subsamples with
different thresholds in stellar mass, $M_{\star}$, and B-band absolute
magnitude, $M_{\rm B}$.  Specifically, we considered the same
subsamples analysed by M13, with four different thresholds in stellar
mass ($\log(M_\star[h^{-2}\,M_\odot])>9$, $9.5$, $10$, $10.5$) and
five in absolute B magnitude ($M_{\rm B}<-19.5$, $-20$, $-20.5$,
$-21$, $-21.5$). The properties of the various samples are
reported in Tables \ref{tab:table1} and \ref{tab:table2}.

\begin{table}[b!]
\small
  \begin{center}
    \caption[]{Properties of the selected VIPERS sub-samples in
      threshold luminosity bins. Absolute magnitudes are quoted in
      units of $M_B(z=1.1)-5\log(h)$.}
    \begin{tabular}{ccccc}
      \hline
      \hline
      redshift & median & magnitude & median  & $N_{gal}$\\
      range & redshift &  range & magnitude &\\
      \hline
      \hline
      $[0.5,0.7]$ & $0.62$ & $<-19.5$ & $-19.87$ & $17473$\\
      $[0.5,0.7]$ & $0.62$ & $<-20.0$ & $-20.15$ & $12432$\\
      $[0.5,0.7]$ & $0.62$ & $<-20.5$ & $-20.49$ & $7472$\\
      $[0.5,0.7]$ & $0.62$ & $<-21.0$ & $-20.86$ & $3599$\\
      $[0.5,0.7]$ & $0.62$ & $<-21.5$ & $-21.28$ & $1236$\\
      \hline
      $[0.7,0.9]$ & $0.79$ & $<-20.0$ & $-20.41$ & $14442$\\
      $[0.7,0.9]$ & $0.80$ & $<-20.5$ & $-20.68$ & $9469$\\
      $[0.7,0.9]$ & $0.80$ & $<-21.0$ & $-21.05$ & $4605$\\
      $[0.7,0.9]$ & $0.80$ & $<-21.5$ & $-21.45$ & $1619$\\
      \hline
      $[0.9,1.1]$ & $0.97$ & $<-20.5$ & $-21.00$ & $5207$\\
      $[0.9,1.1]$ & $0.98$ & $<-21.0$ & $-21.25$ & $3477$\\
      $[0.9,1.1]$ & $0.99$ & $<-21.5$ & $-21.65$ & $1409$\\
      \hline
      \hline
      \label{tab:table1}
    \end{tabular}
  \end{center}
\end{table}
\begin{table}[t!]
\small
  \begin{center}
    \caption[]{Properties of the selected VIPERS sub-samples in
      threshold stellar mass bins. Stellar masses are quoted in units of 
      $[h^{-2}\,M_\star]$.}
    \begin{tabular}{ccccc}
      \hline
      \hline
      redshift & median & stellar mass  & median  & $N_{gal}$\\
      range & redshift & range  & stellar mass & \\
      \hline
      \hline
      $[0.5,0.7]$ & $0.61$ & $>9.0$ & $9.82$ & $17102$\\
      $[0.5,0.7]$ & $0.62$ & $>9.5$ & $10.11$ & $11567$\\
      $[0.5,0.7]$ & $0.62$ & $>10.0$ & $10.35$ & $6880$\\
      $[0.5,0.7]$ & $0.62$ & $>10.5$ & $10.66$ & $2151$\\
      \hline
      $[0.7,0.9]$ & $0.78$ & $>9.0$ & $9.93$ & $15020$\\
      $[0.7,0.9]$ & $0.79$ & $>9.5$ & $10.15$ & $11346$\\
      $[0.7,0.9]$ & $0.79$ & $>10.0$ & $10.39$ & $6884$\\
      $[0.7,0.9]$ & $0.79$ & $>10.5$ & $10.67$ & $2498$\\
      \hline
      $[0.9,1.1]$ & $0.97$ & $>9.5$ & $10.19$ & $4558$\\
      $[0.9,1.1]$ & $0.97$ & $>10.0$ & $10.46$ & $2857$\\
      $[0.9,1.1]$ & $0.97$ & $>10.5$ & $10.71$ & $1281$\\
      \hline
      \hline
      \label{tab:table2}
    \end{tabular}
  \end{center}
\end{table}

The luminosity-redshift and stellar mass-redshift relations for VIPERS
PDR-1 galaxies are shown in Fig. \ref{fig:sub_samples}. Flat
thresholds in stellar mass have been considered, since many works
confirmed a negligible evolution in $M_{\star}$ up to
$z\sim1$\citep{pozzetti2007, pozzetti2010, davidzon2013}.  On the
contrary, the adopted thresholds in absolute magnitude have been
constructed to follow the redshift evolution of galaxies, considering
$M_{\rm B}(z)=M_{\rm B}(0)+z$ \citep{ilbert2005, zucca2009,
  meneux2009,fritz2014}. Figure \ref{fig:sub_samples} shows also the
$90\%$ completeness limits for different galaxy types. As can be
noted, all the magnitude-selected subsamples are volume-limited,
except the ones in the highest redshift bins, where the reddest
galaxies start to fall out of the VIPERS sample. On the other hand,
mass incompleteness affects the subsamples selected in stellar
mass. M13 performed a detailed analysis to quantitatively estimate
this effect, finding a scale-dependent reduction of clustering that
mainly affects very small scales ($\lesssim$1 \Mpch), not significant
for our analysis.


\section{The three-point correlation function}
\label{sec:method}

\subsection{Theoretical setup}
The three-point correlation function estimates the
probability of finding triplets of objects at relative comoving
distances $r_{12}$, $r_{13}$, and $r_{23}$ \citep{peebles1980}. If we
define $\bar{n}$ as the average density of objects, $V_i$ as the
comoving volumes at $\overrightarrow{r_{i}}$, and $\xi$ as the two-point correlation function,
this probability can be written as:
\begin{multline}
d{\rm
  P}=\bar{n}^{3}[1+\xi(r_{12})+\xi(r_{13})+\xi(r_{23})+\zeta(r_{12},r_{13},r_{23})]
\\ dV_{1}dV_{2}dV_{3} \, .
\end{multline}

From the connected 3PCF $\zeta$ it is possible to define also the reduced 3PCF as follows:
\begin{equation}
Q(r_{12},r_{13},r_{23}) \equiv
\frac{\zeta(r_{12},r_{13},r_{23})}{\xi(r_{12})\xi(r_{13})+\xi(r_{13})\xi(r_{23})+\xi(r_{23})\xi(r_{13})}
\, .
\label{eq:3PCF}
\end{equation}
This function, introduced by \citet{groth1977}, has the advantage of
having a smaller range of variation with respect to $\xi$ and $\zeta$,
since it can be shown that in hierarchical scenarios
$\zeta\propto\xi^{2}$ \citep{peebles1975}. Moreover it depends solely on
the bias parameters, and not on $\sigma_{8}$.

\begin{figure}
  \includegraphics[width=0.49\textwidth]{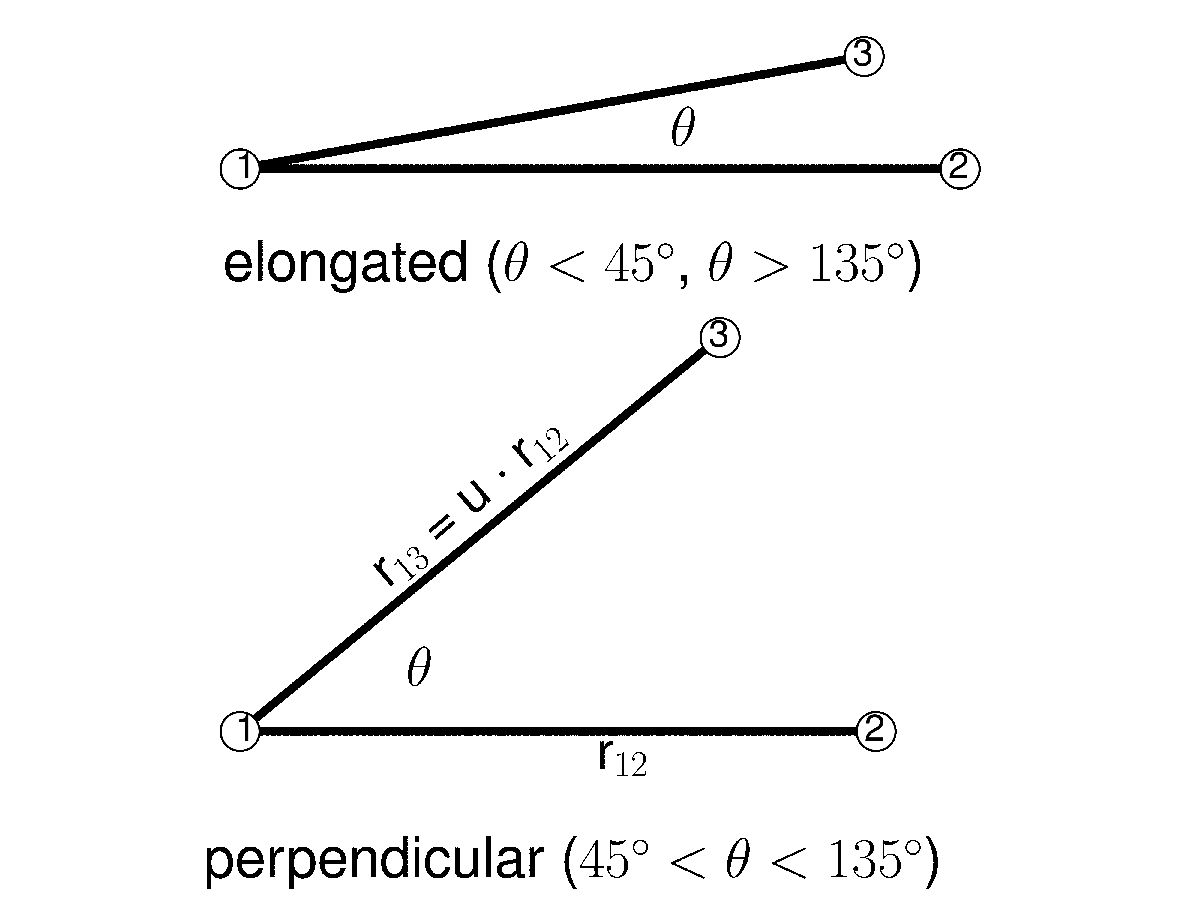}
  \caption{Adopted parameterisation for estimating the 3PCF. 
      Both the perpendicular and elongated configurations are shown,
      for illustrative purposes.}
  \label{fig:3PCF_param}
\end{figure}

Different possible parameterisations have been discussed in literature
to define the triangles, in order to investigate the shape dependence
of the 3PCF \citep{jing1995, gaztanaga2005, nichol2006, kulkarni2007,
  guo2014}.  In this paper, we adopt the parameterisation introduced
by \citet{marin2011}, in which the relation between two sides of the
triangles is fixed, that is $r_{13}\equiv u\cdot r_{12}$, and then the 3PCF
is estimated as a function of the angle $\theta$ between the two
sides:
\begin{equation}
  \left\{
    \begin{array}{l}
      r_{12}\\
      r_{13}\equiv u\cdot r_{12}\\
      r_{23}\equiv r_{12}\cdot\sqrt{1+u^{2}-2\cdot u \cdot \cos\theta}\, . \nonumber 
    \end{array}
    \right.
\end{equation}
In this way, elongated configurations are represented by
$\theta\sim0$ and $\theta\sim\pi$, while perpendicular
configurations by $\theta\sim\pi/2$ (see
Fig. \ref{fig:3PCF_param}). Differently from other parameterisations,
in which for example all the triangle sides are fixed and the 3PCF is only
measured as a function of scale, the adopted configuration is
particularly convenient to study at the same time the scale dependence
of the 3PCF, by changing the length and the ratios between the first
two triangle sides, and the shape dependence, as a function of the
angle $\theta$.  As in \citet{marin2011}, we consider a constant
logarithmic binning in $\Delta r_{ij}/r_{ij}$. It has been
demonstrated that this binning scheme allows one to include in each
$\theta$-bin triangles with similar shapes, providing also smaller
errors with respect to other parameterisations.

\subsection{Estimator and implementation}
The \citet{szapudi1998} estimator is used to measure the 3PCF:
\begin{equation}
\zeta(r_{12},u\cdot r_{12},\theta) = \frac{DDD-3DDR+3DRR-RRR}{RRR} \, ,
\end{equation}
where $DDD$, $RRR$, $DDR$, and $DRR$ are the normalised numbers of
data triplets, random triplets, data-data-random triplets, and
data-random-random triplets respectively.  The 2PCF is measured with
the \citet{landy1993} estimator:
\begin{equation}
\xi(r) = \frac{DD-2DR+RR}{RR} \, ,
\end{equation}
where $DD$, $RR,$ and $DR$ are the normalised numbers of data pairs,
random pairs, and data-random pairs respectively.

To measure both the 2PCF and the 3PCF we exploit the
{\small CosmoBolognaLib}, a large suite of \texttt{C++} libraries for
cosmological calculations \citep{cosmobolognalib} \footnote{The
  {\small CosmoBolognaLib} can be freely downloaded at
  https://github.com/federicomarulli/CosmoBolognaLib.}. Pair and
triplet counts are computed with a chain-mesh algorithm, that
allows us to significantly reduce the computing time by optimizing the search
in the surveyed volume. Specifically, as a first step of the procedure, both
the data and random catalogues are pixelized, that is they are divided
into small sub-regions, and the indices of the objects belonging to each
sub-region are stored in vectors. Then, the count is computed by
running the algorithm only on the sub-regions actually contributing to
the pair and triplet counts in the chosen scale range.

The random sample for each stellar mass and luminosity bin has
been created implementing the same observing strategy of real data,
with $\rm N_R=60 N_{gal}$, where $N_R$ is the number of random objects
and $N_{gal}$ is the number of galaxies in each sample.  The redshifts
of the random objects are drawn by the observed radial distribution
of the W1+W4 VIPERS samples, conveniently smoothed as described in M13.


\begin{figure*}
  \includegraphics[width=0.49\textwidth]{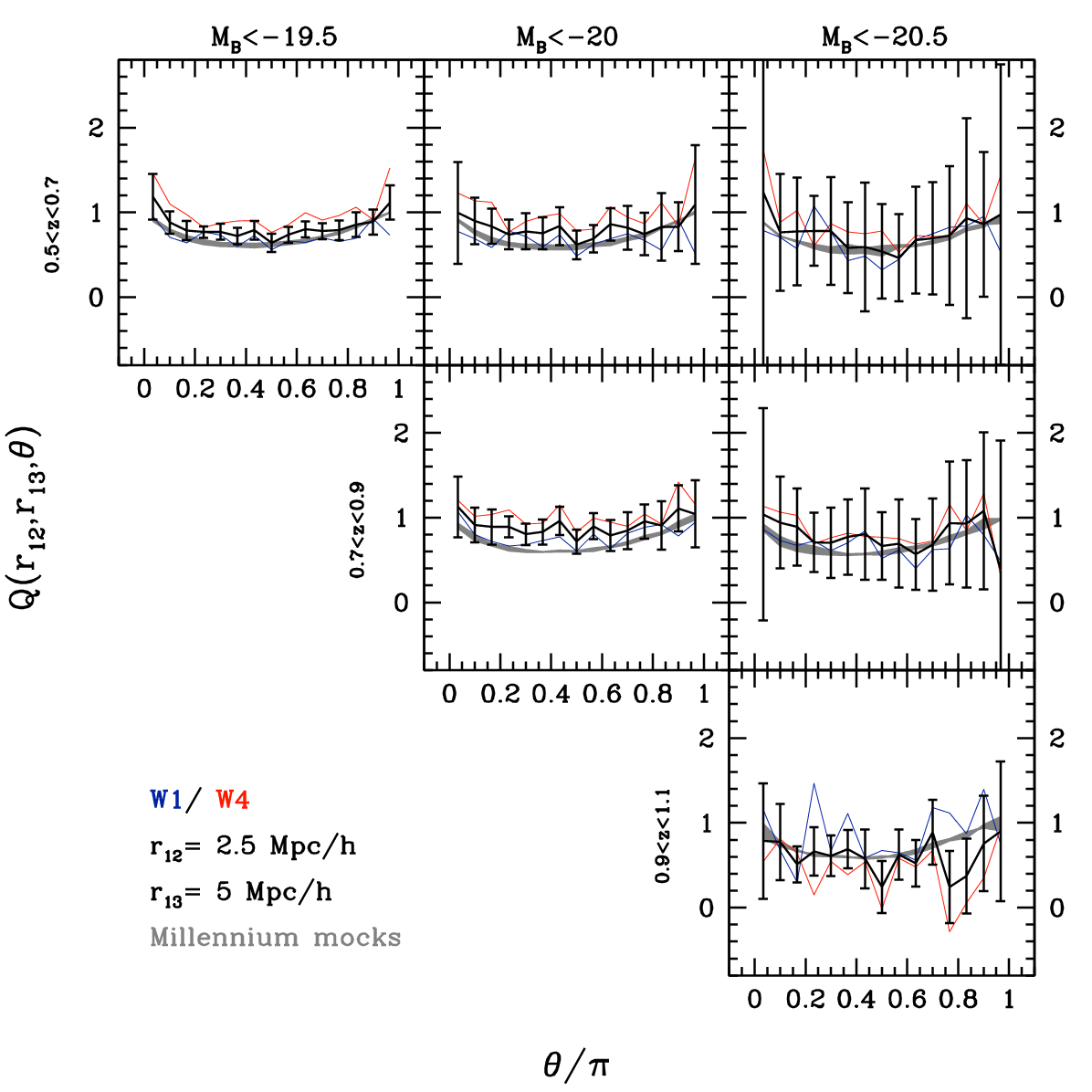}
  \includegraphics[width=0.49\textwidth]{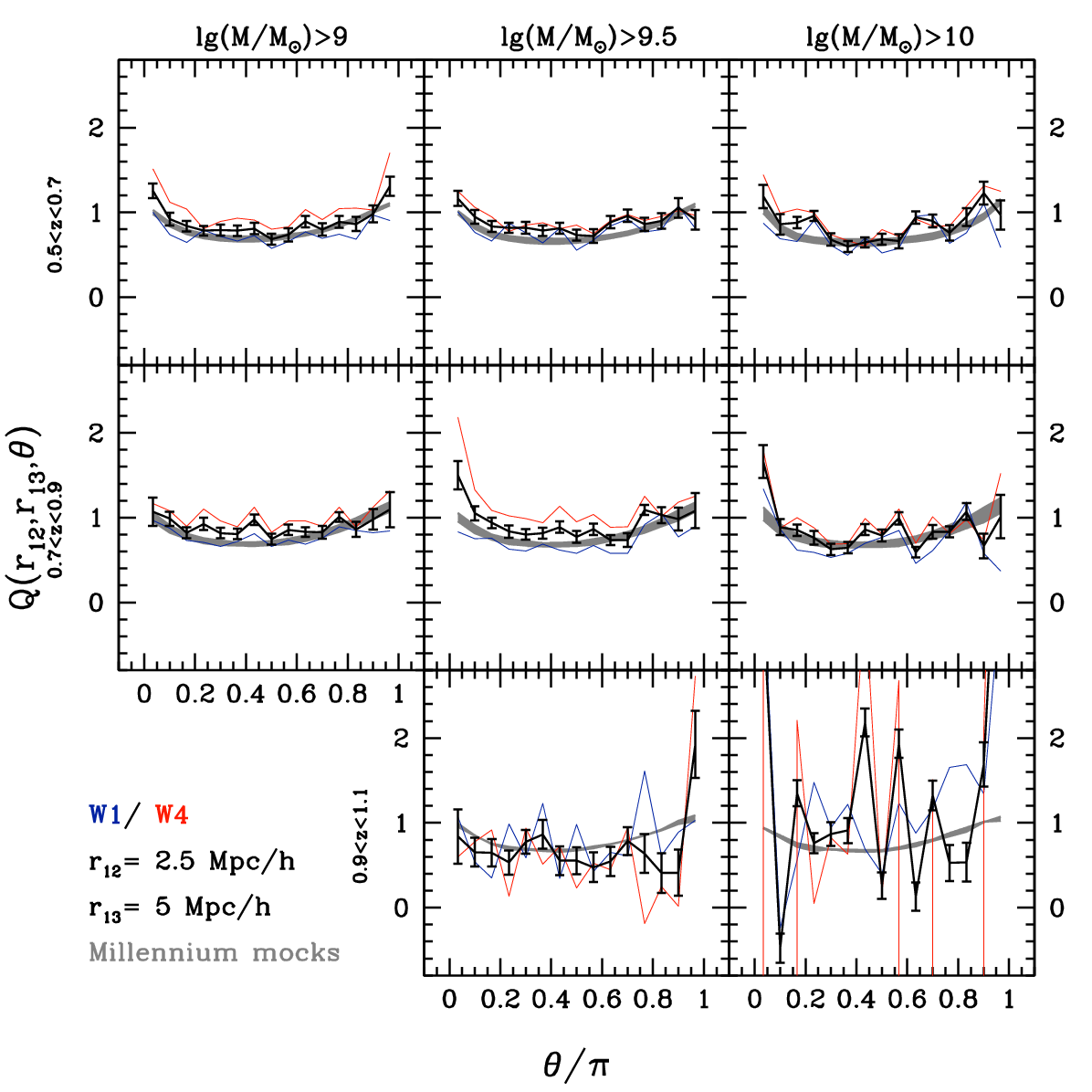}
  \caption{Redshift-space reduced 3PCF, $Q(\theta)$, as a function of
    redshift (panels from top to bottom) for different luminosity
    (left plots) and stellar mass thresholds (right plots), for scales
    $r_{12}=2.5$ \Mpch and $r_{13}=5$ \Mpch. Blue and red lines show
    the measurements in the W1 and W4 fields, respectively, while the
    black lines show the combined 3PCF, with its associated errors.
    The grey shaded area shows the redshift-space 3PCF measured in two
    large Millennium mock catalogues, constructed to mimic the
    properties of VIPERS data, and is not representative of the
    expected scatter in the data.}
  \label{fig:3PCF_2.5}
\end{figure*}

\subsection{Error and weight estimate}
\label{sub:errors}

We estimate the errors on the measured 3PCF using the set of 26 mock
galaxy catalogues used in M13, and described in \cite{delatorre2013b}.
Halo occupation distribution (HOD) mocks calibrated on the real data
are used for the luminosity-selected samples, while mocks implementing
the stellar-to-halo mass relation (SHMR) of \citet{moster2013} are
used for the stellar mass-selected samples \citep{delatorre2013b}. In
both cases, galaxies are assigned to dark matter haloes extracted from
the MultiDark N-body simulation \citep{prada2012}. To populate the
simulation with haloes below the mass resolution limit, in order to
reach the haloes hosting the very faint VIPERS galaxies, we exploit
the technique described in \citet{delatorre2013a}. These mocks
  are not optimised to reproduce higher-order
  correlations. Nevertheless, they provide a fair representation of
  the data, with a reduced 3PCF in good agreement with our
  measurements in all the subsamples analysed.

The covariance error matrix is estimated from
the dispersion among the mock catalogues:
\begin{equation}
\displaystyle C_{ij} =
\frac{1}{N}\sum_{k=1}^{N-1}\left(Q_{i}^{k}-\bar{Q}_{i}\right)
\left(Q_{j}^{k}-\bar{Q}_{j}\right) \, ,
\end{equation}
where $\bar{Q}_{j}$ is the mean value of the reduced 3PCF averaged between
the 26 mock catalogues.  The errors on the 3PCF are then obtained from
the square root of the diagonal values only of the covariance matrix,
$\sigma_i=\sqrt{C_{ii}}$, applying on mock subsamples the selection
criteria used on real data. Analogously, also the covariance error matrix for
the connected 3PCF is estimated. Further information regarding the covariance
matrices can be found in Appendix \ref{appendix}. Given the limited number of 
mocks available, we consider only the diagonal elements of $C_{ij}$, and discuss
 in the Appendix the effect of considering the full covariance.

Following the same procedure described in M13, we apply to each galaxy
a weight that depends on both its redshift, $z$, and position in the quadrant, Quad, given by:
\begin{equation}
w({\rm Quad},z) = w_{TSR}({\rm Quad})\cdot w_{SSR}({\rm Quad})\cdot w_{CSR}(z) \nonumber ,
\label{eq:weight}
\end{equation}
and that accounts for three independent sources of systematic errors, each one quantified
by its own weight: 
\begin{itemize}
\item $w_{TSR}$ is the weight that accounts for the
target sampling rate, that is the probability that a galaxy
in the photometric catalogue has a spectroscopic redshift measurement,
and it is computed as the ratio between the total number of galaxies
in the photometric catalogue and the ones actually spectroscopically
targeted;
\item $w_{SSR}$ is the weight that accounts for the 
spectroscopic success rate, that is the probability that 
a galaxy spectroscopically targeted has a reliable
redshift measurement, that is a redshift with flag $2\leq z_{\rm
flag}\leq9.5$ \citep[for a more detailed discussion on VIPERS
redshift flags, we refer to][]{guzzo2014};
\item $w_{CSR}(z)$ is the weight that accounts for the
colour sampling rate, defined as CSR$(z)=0.5[1-{\rm erf}(7.405-17.465\cdot z)]$, 
and taking into account the incompleteness due to the VIPERS colour selection.
\end{itemize}
Both  $w_{TSR}$  and  $w_{SSR}$ mainly depend on the quadrant 
${\rm Quad}$ and on the redshift $z$, with all other possible dependencies 
being negligible \citep{delatorre2013b}. 
We verified that these weights do not significantly affect our
results, typically changing the clustering measurements below the
estimated $1\sigma$ uncertainties.


\section{Results}
\label{sec:results}

In this section, we present the measurements of the redshift-space 3PCF for the
VIPERS sample, discussing their dependence on shape, redshift, stellar
mass, and luminosity in comparison with the results found in the literature. The
analysis has been performed both in W1 and W4 separately and
combining the counts in the two fields to provide a single
measurement. The 3PCF has been measured in various redshift, stellar
mass, and luminosity bins, as discussed in Sect. \ref{sec:data}.
Differently from M13, we will show and discuss here only the results
for the three lowest stellar mass and luminosity samples, since the
most extreme bins result to be noise-dominated (namely $M_{\rm
  B}<-21$, $-21.5$ and $\log(M_\star[h^{-2}\,M_\odot)>10.5$).

To explore the dependence of the clustering also on the scales and
shapes of galaxy triplets, we analyse three different scales, with
$r_{12}=2.5,5,10$ \Mpch. Throughout this
analysis, we consider a ratio between the first and the second side of the
triplet $r_{13}=2\cdot r_{12}$, and for each of
these configurations we use $15$ equi-spaced angular bins in $\theta$.
The high galaxy number density in VIPERS allows
us to explore the 3PCF down to scales smaller than other surveys
at similar redshift \citep[e.g. WiggleZ,][]{marin2013}. The choice of $r_{13}/ r_{12}=2$
is justified to avoid having strongly non-linear configurations that would appear in collapsed triangles
for $r_{13}/ r_{12}=1$, and to allow comparison with similar analyses in the literature
\citep{marin2011,mcbride2011,marin2013}.


\subsection{Redshift and scale dependence}
\label{sec:redendence}

\begin{figure*}
  \includegraphics[width=0.49\textwidth]{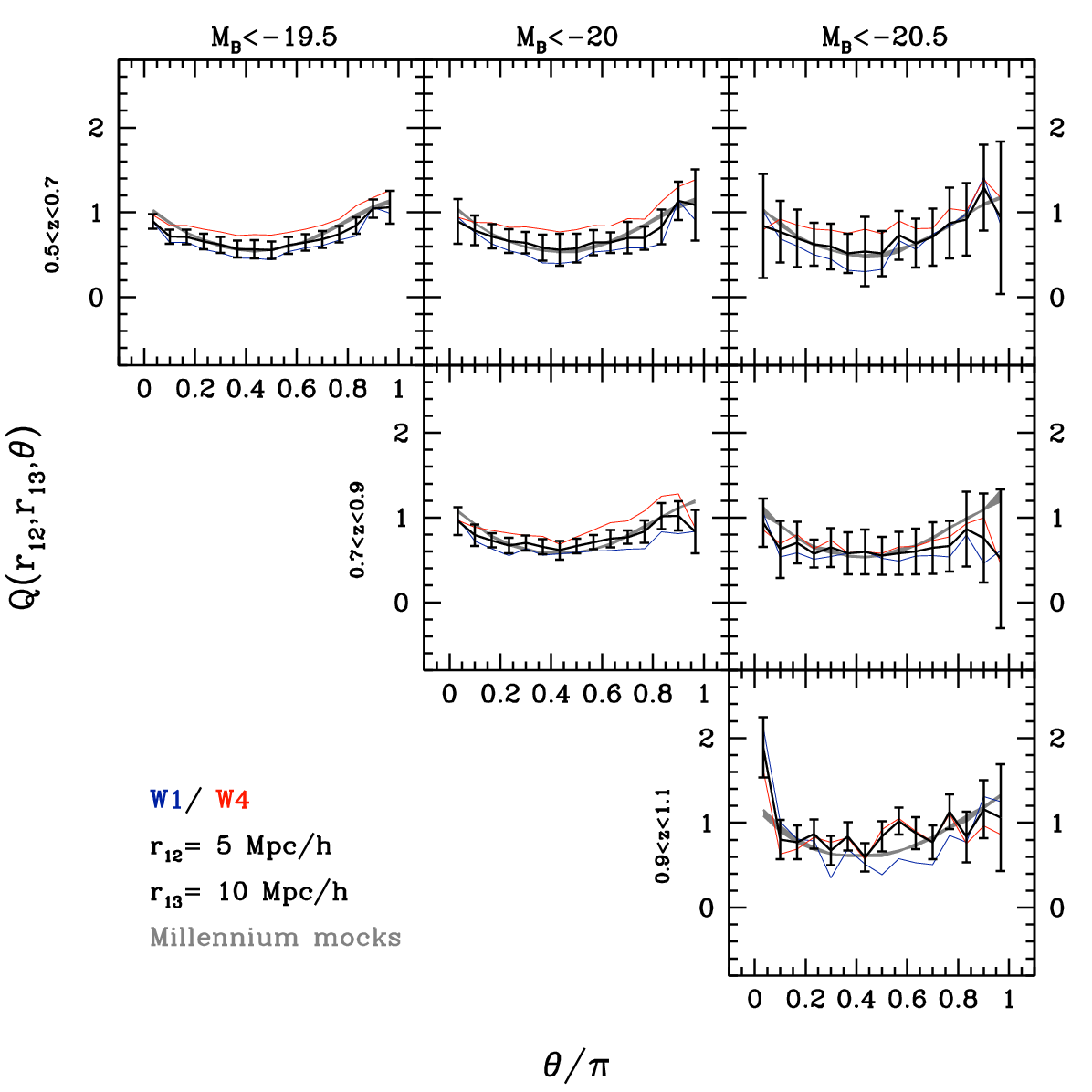}
  \includegraphics[width=0.49\textwidth]{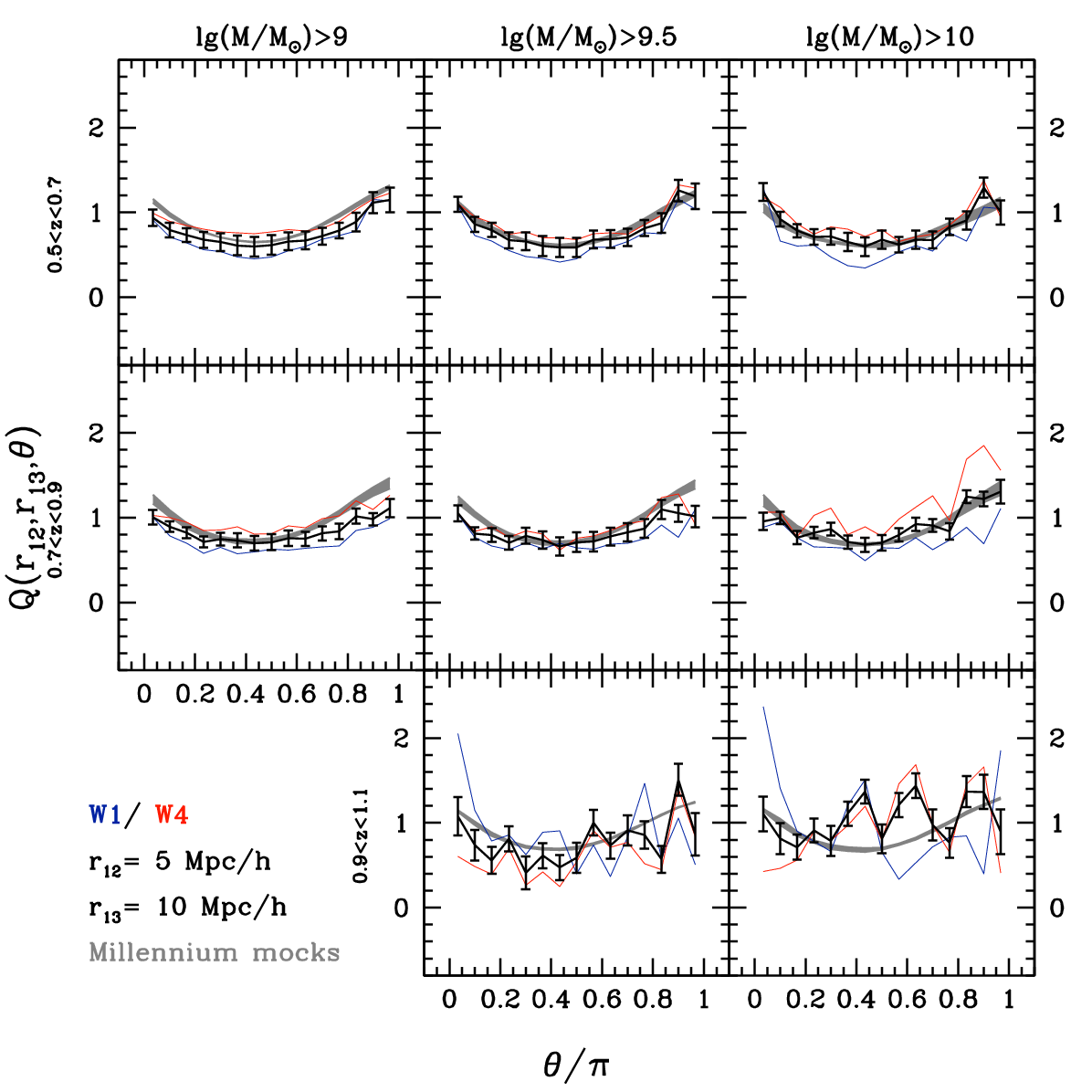}
  \caption{Same as Fig. \ref{fig:3PCF_2.5}, but for scales $r_{12}=5$ \Mpch and $r_{13}=10$ \Mpch.}
  \label{fig:3PCF_5}
\end{figure*}
\begin{figure*}
  \includegraphics[width=0.49\textwidth]{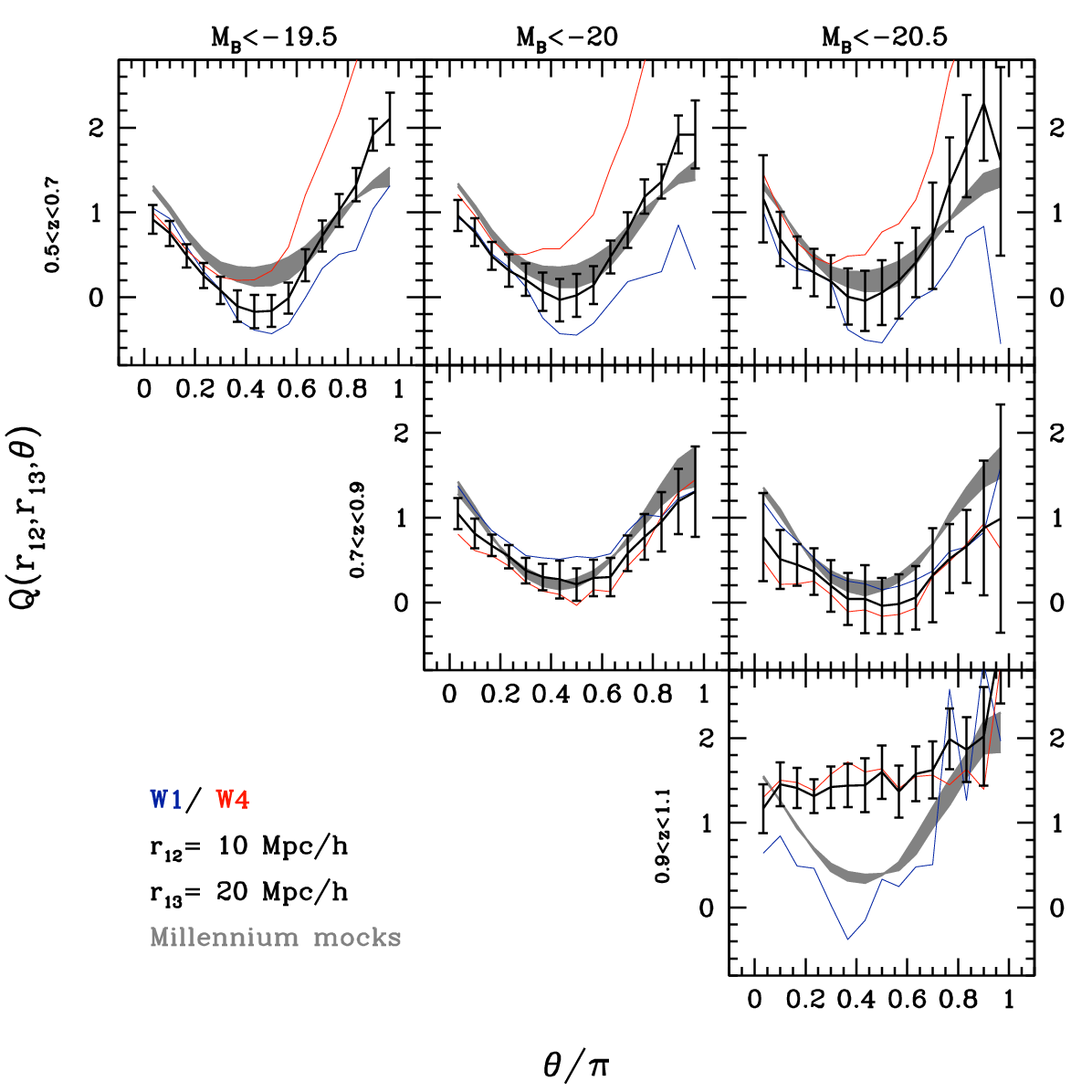}
  \includegraphics[width=0.49\textwidth]{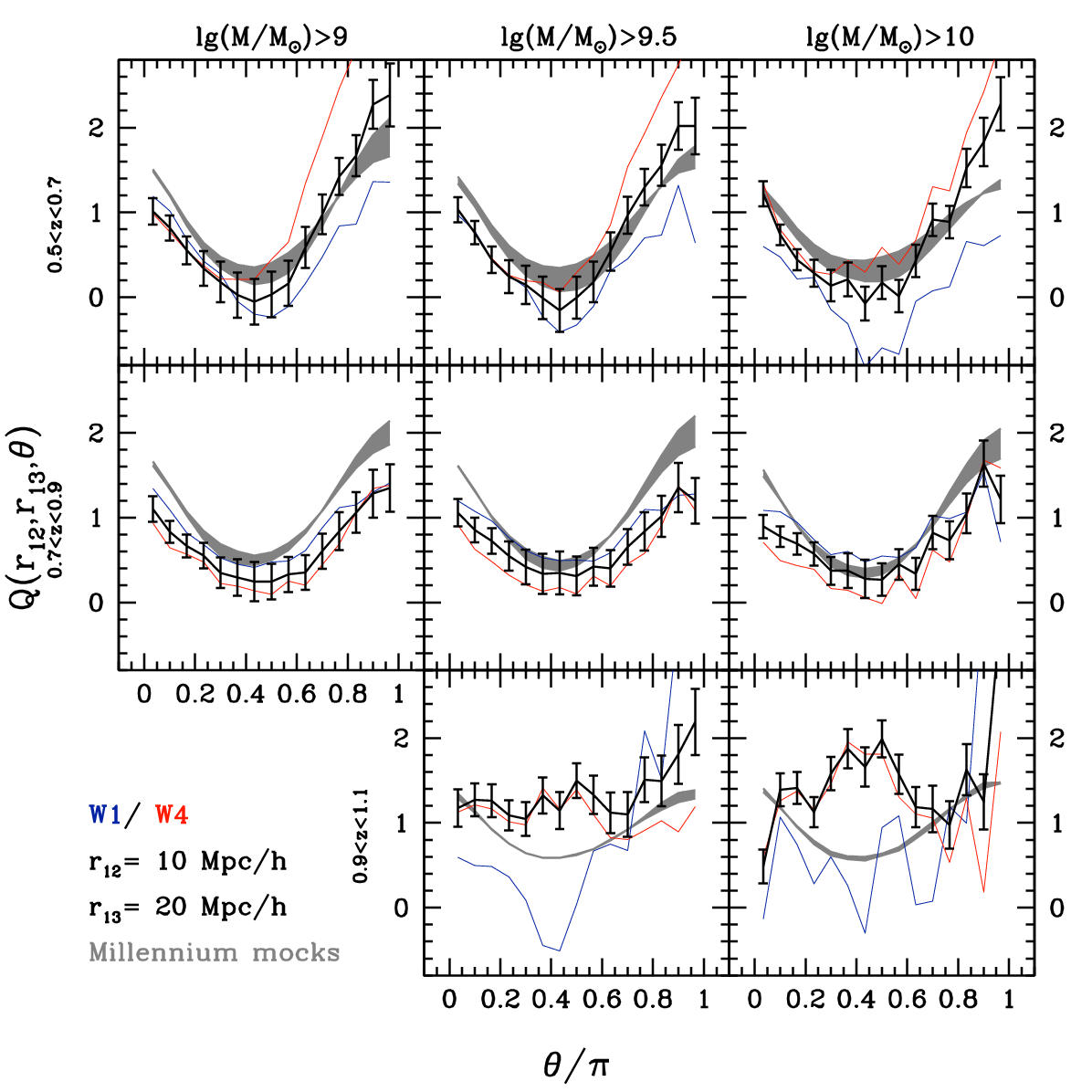}
  \caption{Same as Fig. \ref{fig:3PCF_2.5}, but for scales $r_{12}=10$ \Mpch and $r_{13}=20$ \Mpch.}
  \label{fig:3PCF_10}
\end{figure*}

Figures \ref{fig:3PCF_2.5}, \ref{fig:3PCF_5}, and \ref{fig:3PCF_10} show
the redshift-space reduced 3PCF as a function of luminosity and
stellar mass, in three redshift ranges and for three different
configurations, for $r_{12}=2.5,5,10$ \Mpch respectively. Both the 3PCF of W1 and W4
fields and the combined 3PCF are shown.

This analysis is in qualitative agreement with many previous works performed on both
numerical simulations \citep{gaztanaga2005, marin2008, moresco2014}
and real data \citep{mcbride2011, marin2011, marin2013, guo2014}, 
where a transition is found from the ``U'' shape of $Q(\theta)$ to the ``V'' shape moving from the smaller
to the larger scales. This feature indicates a more pronounced anisotropy in $Q(\theta)$
at increasing scales which is theoretically expected. It is related to the
different shapes of structures at different scales, with compact, spherically symmetric
structures dominating at small distances, and filamentary structures
starting to contribute at larger scales, as indicated by the larger
value of $Q(\theta)$ in the elongated configurations.

At small scales no significant evolution with redshift is
found. However, in these configurations $Q(\theta)$ is generally
flatter, and differences are harder to detect. On the contrary,
at larger scales it is possible to see a clear trend, with the 3PCF
being flatter at higher redshifts. This can be interpreted as an
indication of the build-up of filaments with cosmic time, that evolve
enhancing the 3PCF in elongated configurations while reducing it in
the equilateral configurations.

The differences between the 3PCF measurements in W1 and W4 are
  caused by small density fluctuations on large scales similar to those of the two samples. Similar
  trends at the scales probed by our analysis are also found in the
  2PCF, in agreement with what is also found in our mock
  catalogues. The covariance between different scale bins makes these
  differences systematic. Moreover, the chosen 3PCF configurations are
  not independent, being partially overlapping. Thus, the trends found
  in different configurations are partially correlated. 

\begin{figure*}
\begin{center}
  \includegraphics[angle=-90,width=0.9\textwidth]{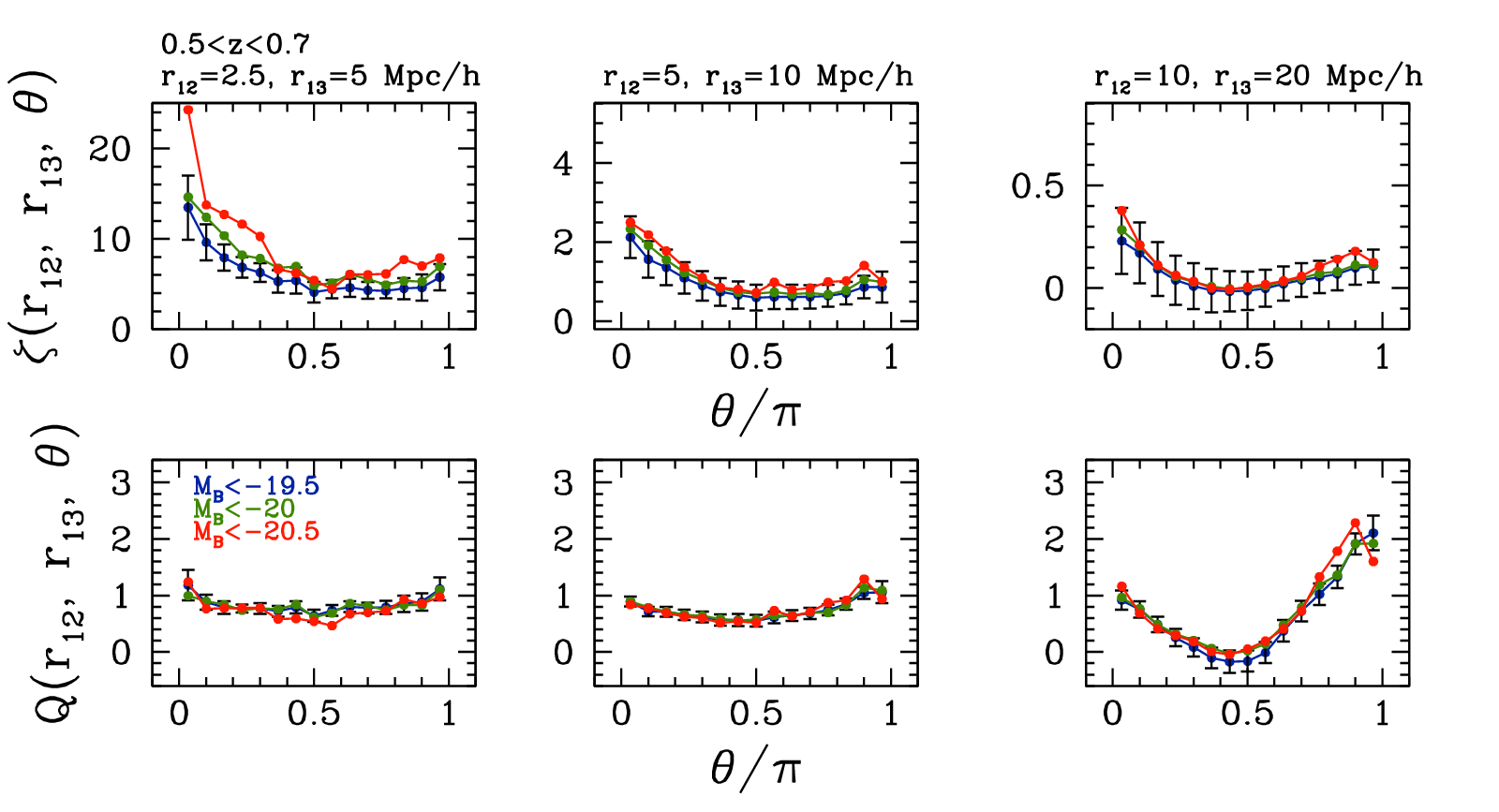}
  \caption{Redshift-space connected 3PCF (upper panels) and reduced
    3PCF (lower panels) as a function of luminosity 
    at redshift $0.5<z<0.7$ for different
    scales. Different colours show the measurements in the lower
    (blue), intermediate (green), and higher (red) threshold bins. For
    clarity reason, error-bars are shown only in the lower bin.}
  \label{fig:3PCF_dependence1}
\end{center}
\end{figure*}

\subsection{Luminosity and stellar mass dependence}
\label{sec:dependence}

\begin{figure*}
\begin{center}
  \includegraphics[angle=-90,width=0.9\textwidth]{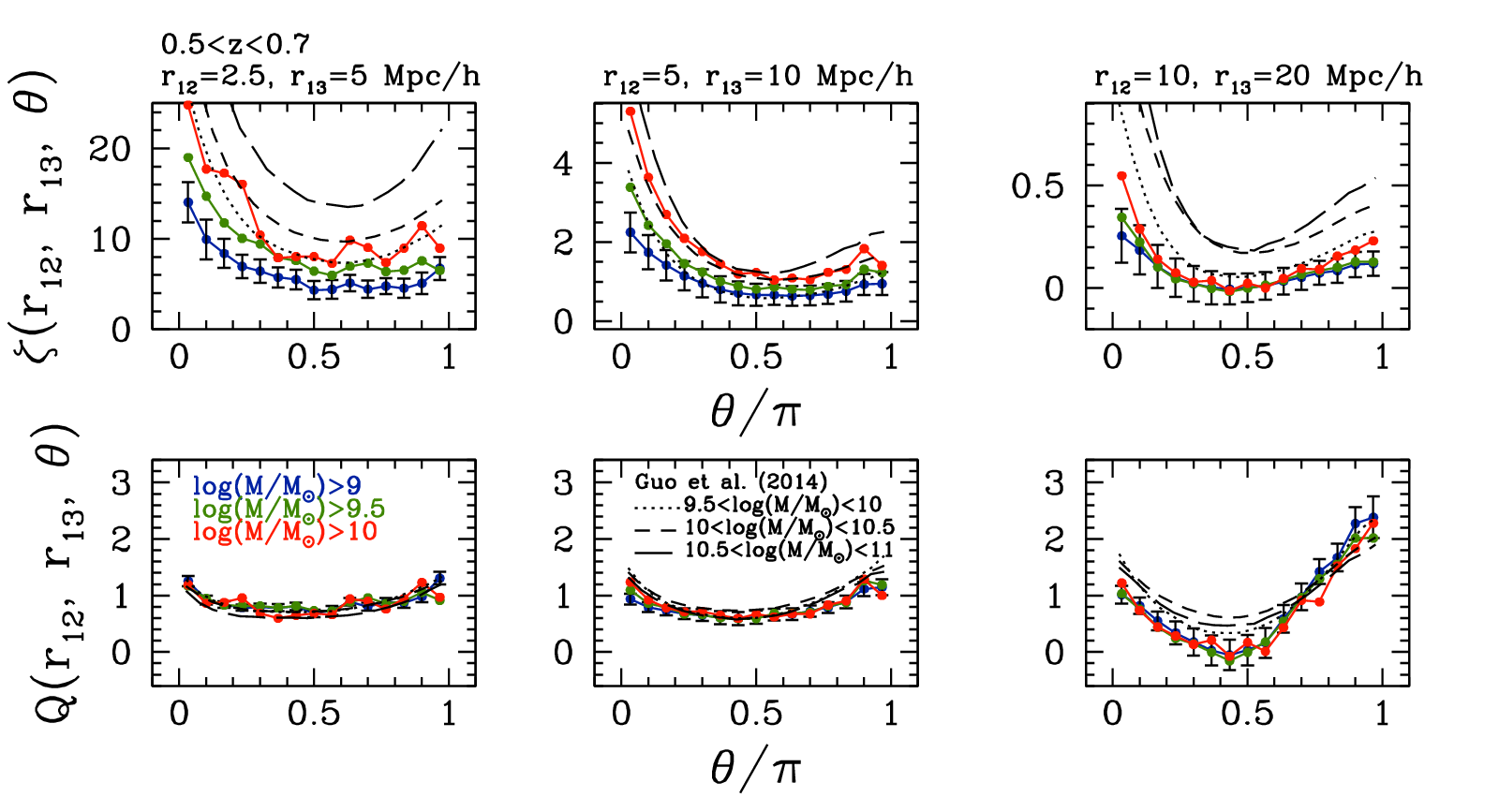}
  \caption{Same as Fig. \ref{fig:3PCF_dependence1}, but as a function
    of stellar mass.  As a comparison, the 3PCF measurements
      obtained by \citet{guo2014} on SDSS-DR7 main sample are also
      shown, for different stellar mass selected galaxy catalogues.
      Dotted, dashed, and solid lines represent subsamples with
      $9.5<\log(M/M_\odot])<10$, $10<\log(M/M_\odot])<10.5$, and
  $10.5<\log(M/M_\odot])<11$, respectively.}
  \label{fig:3PCF_dependence2}
\end{center}
\end{figure*}

In Figs. \ref{fig:3PCF_dependence1} and \ref{fig:3PCF_dependence2} we show the luminosity and
stellar mass dependence in redshift space for both the connected and the
reduced 3PCF, at the various scales explored in this analysis. In
particular, we focus on the lower redshift range, $0.5<z<0.7$, where
we have a larger leverage in observed properties.

In general, the connected 3PCF $\zeta(\theta)$ exhibits a much
stronger dependence on both stellar mass and luminosity than the
reduced 3PCF $Q(\theta)$, particularly evident at small scales.  
  We find an average difference of $\sim$40\% in $\zeta(\theta)$
  between the higher and lower mass bins, and of only $\sim$5\% in
  $Q(\theta)$ in the same bins. Similarly, we have a difference of
  $\sim$20\% and of $\sim$0-10\% between the higher and the lower
  magnitude bins in $\zeta(\theta)$ and $Q(\theta)$,
  respectively. These differences, however, are not statistically
  significant considering the estimated uncertainties of the present
  analysis.

This differential trend in $Q$ and $\zeta$ agrees with that found by
\citet{guo2014} from the analysis of a more local sample extracted
from the Sloan Digital Sky Survey - Data Release 7 (SDSS-DR7).
 For a more direct comparison, we report their
  measurements in Fig. \ref{fig:3PCF_dependence2}. We note that while
  this analysis adopted the same configuration used in this paper
  ($r_{13}/ r_{12}=2$), an homogeneous comparison is not possible, due
  to the different binnings. Moreover, given the different choices of
  absolute magnitude thresholds (the considered r-band absolute
  magnitude limits), we decided to compare only the results obtained
  in stellar mass selected samples. In particular, they considered
  $r_{13}\sim2.5, 5.5, 9,$ and $9.5<\log(M/M_\odot])<10$,
$10<\log(M/M_\odot])<10.5$, and $10.5<\log(M/M_\odot])<11$, and found
    an average difference of $\sim$8\% between the higher and lower
    mass bins in $Q(\theta)$, and of $\sim50$\% in $\zeta(\theta)$.

This effect can be explained considering
the different sensitivity of the two functions on the bias
parameters. According to second-order perturbation theory and in the
hypothesis of linear bias \citep{fry1993, frieman1994}, the 2PCF of dark matter can be connected
to the one of galaxies through the bias parameter $b_{1}$, with a
relation $\xi_{g}=b_{1}^{2}\xi_{m}$; similarly, for the 3PCF it is
possible to derive the relation $\zeta_{g}\sim b_{1}^{3}\zeta_{m}$,
that is valid at the first order.  Hence, from Eq.~\ref{eq:3PCF} we have that
$Q\propto1/b_{1}$, and therefore the dependence of $\zeta$ on the linear bias
is more significant. While our analysis finds a detectable trend in $\zeta$, $Q$ shows
no significant trend either in stellar mass or in luminosity, in contrast with some 
earlier results that find a slight dependence with luminous and massive galaxies 
having lower amplitudes of $Q$ \citep{jing2004, mcbride2011, guo2014}.  
The absence of any significant trend in our measurements could be due to the different
redshift ranges probed. While most previous works focused on local
galaxies ($z<0.3$), our analysis has been performed at $z>0.5$, 
where the amplitude of the clustering and the non-Gaussianity due 
to non-linear evolution is smaller, reducing the differences in $Q$.
Moreover, the fact that no dependence on luminosity or stellar mass is
detectable in $Q$, despite the presence of a non-zero $b_{1}$ (as shown by
the analysis of the 2PCF on the same dataset, see M13), implies that a non-linear
contribution to the bias should be present, as confirmed by the analysis 
in Sect. \ref{sec:bias}.

From the analysis of the 3PCF it can be found that more luminous and
massive galaxies present a higher clustering, in agreement with the
results obtained on the same sample from the 2PCF by M13.  In
  their analysis they also found a difference of $\sim$15-20\% at the
  same scales probed by our work.  The larger dependence at smaller
scales also confirms the results found by \citet{guo2014} from
the analysis of SDSS-DR7. \citet{cappi2015} also analysed VIPERS data,
measuring volume-averaged higher-order correlation functions. A direct
comparison of the results is not straightforward, since in their
analysis they measured the galaxy normalised skewness $S_{3}$, which,
in hierarchical models, corresponds to $S_{3}\sim3Q$; however, this
function is insensitive to spatial configuration, and any shape
dependence is washed out. Nevertheless, both $S_{3}$ and $Q$ quantify
the contribution to higher-order correlation functions, and similarly
a negligible dependence on luminosity was found.
These results can be interpreted within the hierarchical formation scenario, 
in which more massive and luminous systems
are more clustered than less massive and luminous ones.

\subsection{Comparison with semi-analytic models}
In Figs. \ref{fig:3PCF_2.5}, \ref{fig:3PCF_5}, and \ref{fig:3PCF_10},
we compare our measurements with theoretical
expectations. Specifically, we consider mock galaxy catalogues
constructed on top of the Millennium simulation \citep{springel2005},
by using the Munich semi-analytic model \citep{blaizot2005,
delucia2007}. The same stellar mass and luminosity thresholds used
for real data (see Sect. \ref{sec:data}) have been applied to select
mock galaxy samples.

As can be noted, there is a general agreement between our
measurements and theoretical predictions, except in the higher
redshift bins. Similar results have been found also for the 2PCF
\citep{marulli2013}. The results shown here extend that analysis to
higher-order clustering. Assuming a local bias scenario, the 3PCF of dark matter can be 
connected to the one of galaxies following the equation 
$Q_{g}=1/b_{1}(Q_{dm}+b_{2}/b_{1})$, where $b_{1}$ is the linear
bias parameter and $b_{2}$ the non-linear one.
At the smallest scales (Fig. \ref{fig:3PCF_2.5}), the mocks tend to 
present a smaller normalisation of $Q$ for all redshifts and luminosities,
except in the highest redshift bins, which can be interpreted in theoretical 
models failing to exactly reproduce the non-linear part of the evolution.

At larger scales ($r_{12}$=5 \Mpch, Fig. \ref{fig:3PCF_5}) the agreement with 
data is  better, while at the largest probed scales (Fig. \ref{fig:3PCF_10})
the mocks tend not to reproduce the $\theta$-dependence of $Q$, that may
point to an inconsistency with the observed linear bias. The significance of this discrepancy, however,
might be reduced once the full covariance error matrix is taken into account,
which appears particularly relevant at the larger scales (see Appendix \ref{appendix}).

\subsection{Constraining the galaxy bias}
\label{sec:bias}

\begin{figure}[t!]
\begin{center}
  \includegraphics[angle=0,width=0.48\textwidth]{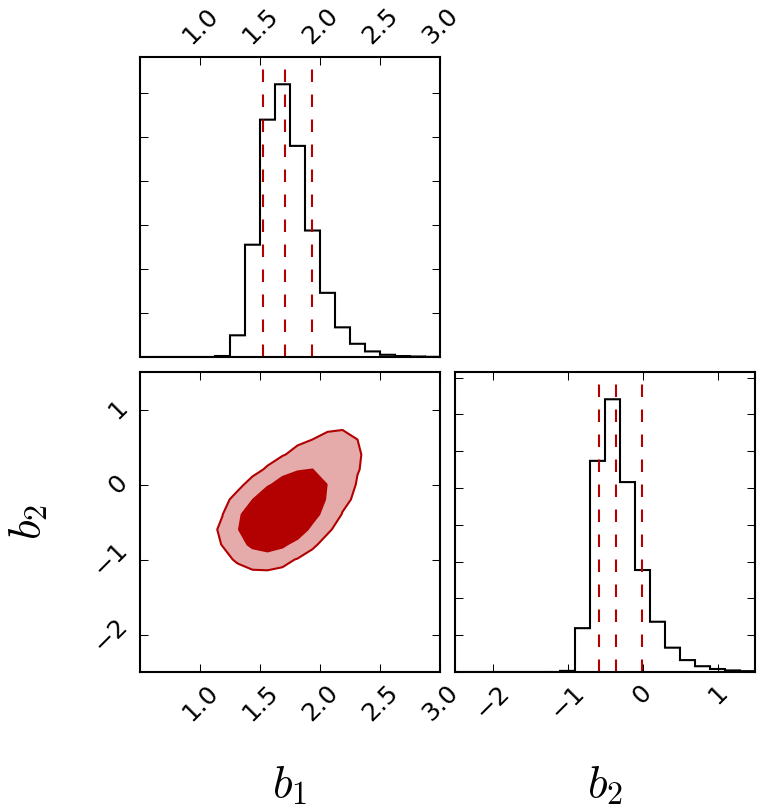}
  \caption{Constraints on $b_1$ and $b_2$ from the
      $\log(M/M_\odot])>9$ sample, at scales $r_{12}=10$ \Mpch and
    $r_{13}=20$ \Mpch. The contour plot shows the $68\%$ and $95\%$
    confidence levels, while the histograms show the posterior
    distributions for the two parameters. The red dashed lines show
    the mean, and the $16^{th}$ and $84^{th}$ percentiles.}
  \label{fig:b1b2}
\end{center}
\end{figure}

\begin{figure*}[ht!]
\begin{center}
  \includegraphics[angle=0,width=0.95\textwidth]{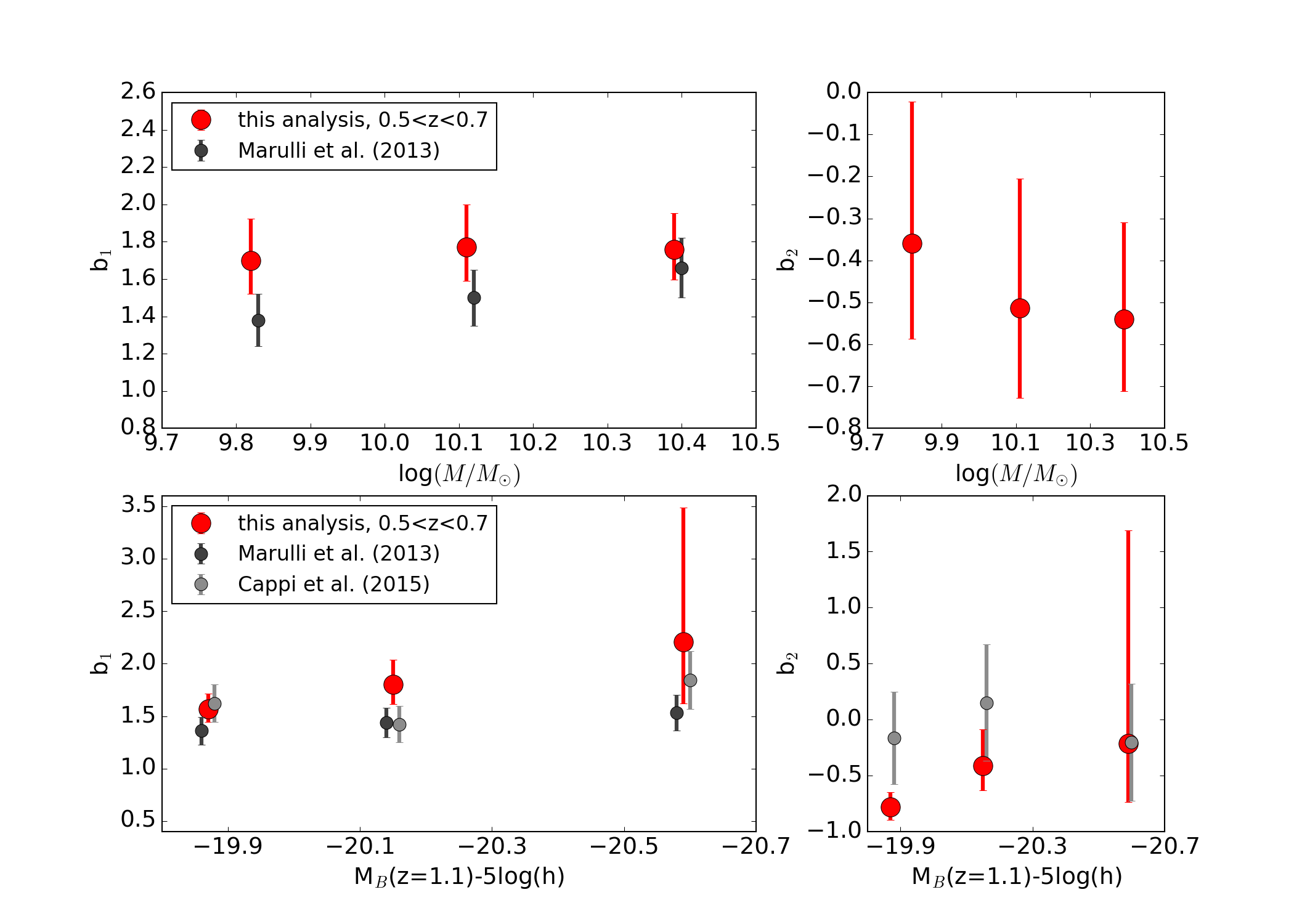}
  \caption{Galaxy linear bias (left panels, $b_1$) and non linear bias
      (right panels, $b_2$) as a function of
      stellar mass (upper panel) and luminosity (lower panel).
      Independent measurements obtained on the same subsamples are
      shown for comparison: from the 2PCF \citep[black
        points][]{marulli2013} and from volume-averaged higher-order
      correlation functions \citep[grey points][]{cappi2015}. The
      values on the x-axis have been slightly offset, for illustrative
      purposes.}
  \label{fig:bias}
\end{center}
\end{figure*}

Finally, we derive constraints on the galaxy bias by using the
  measurements presented in the previous sections. At large scales, it
  is possible to relate the galaxy overdensities to the ones of dark
  matter adopting the local bias model expansion up to the second
  order \citep[see e.g.][]{marin2013,moresco2014,hoffmann2015}, and in
  particular through a linear and a non-linear bias parameter, $b_1$
  and $b_2$, respectively.

One of the most convenient methods to constrain these parameters is to
exploit the reduced 3PCF, $Q_{g}$, since its relation with the reduced 3PCF
of dark matter, $Q_{DM}$, is independent of
$\sigma_8$. The relation between these two quantities in the local
bias model can be expressed as:
\begin{equation}
Q_{\rm g}(r_{12}, r_{13}, \theta)=\frac{1}{b_{1}}\left(Q_{\rm
  DM}(r_{12}, r_{13}, \theta)+\frac{b_{2}}{b_{1}}\right) \, .
\label{eq:bias3PCF}
\end{equation}
We note that in Eq.~\ref{eq:bias3PCF}, $Q_{g}$ represents the
reduced 3PCF in real space, whereas the measured one is in redshift
space.  While the relation between real-space and redshift-space 2PCF
can be modelled fairly accurately \citep{kaiser1987}, this is not the
case for higher-order statistics, especially in configuration
space \citep[however, see][for a model of the effect in Fourier space]{scoccimarro1999}. 
We assessed the impact of this effect using our mocks and found that
the differences between the reduced 3PCF in real and redshift space
are much smaller than the errors associated with the current 3PCF
measurements. Therefore, we neglect dynamic distortions in this
analysis, and use Eq.~\ref{eq:bias3PCF} to model our measurements. A
similar approximation has been used also by \citet{pan2005},
\citet{gaztanaga2005}, and \citet{marin2013}.

To estimate the bias parameters, we consider the largest scales
available in our analysis, namely $r_{12}=10$ \Mpch, probing triangles
whose sides span the range $\rm10<r[$\Mpch$]<30$. The dark matter
reduced 3PCF, $\rm Q_{DM}$, has been measured from the DEMNUni
simulations \citep{carbone2016,castorina2015}, which is a set of large-volume,
high-resolution cosmological N-body simulations, devised in particular
to study the effects of massive neutrinos on the evolution of cosmic
structures. For this work, we considered the $\Lambda$CDM set. These
simulations have much larger volume than the other simulations
considered in this work, allowing us to estimate the dark matter 3PCF
very accurately. We note that the cosmology assumed in this simulation
is slightly different from the one assumed throughout the paper, but, as
also discussed by \cite{marulli2013}, such small differences (especially
in $\Omega_M$) produce an effect much smaller than current errors.

To constrain $b_1$ and $b_2$, we perform a fit using the model given
Eq. \ref{eq:bias3PCF} with a standard $\chi^2$ approach, sampling the
parameter space with the public python package {\tt emcee}
\citep{emcee}, which implements a Markov Chain Monte Carlo
affine-invariant ensemble sampler, as proposed by \citet{goodman2010}.
We focus our analysis in particular on the redshift range $0.5<z<0.7$,
since we verified that at larger redshifts errors are too large to constrain the bias.
An example of best-fit bias constraints obtained is reported in
Fig. \ref{fig:b1b2}, while Fig. \ref{fig:bias} shows the estimated $b_1$
and $b_2$ parameters for both mass and luminosity selected
samples at $0.5<z<0.7$, compared to the results obtained by \citet{marulli2013} from
the 2PCF, and by \citet{cappi2015} from $S_3$, on the same subsamples.

We find that the linear bias parameter, $b_1$, in both stellar mass and luminosity
samples ranges between $1.5<b_1<2$. Our measurements overestimate
slightly $b_1$ with respect to other probes. It is a 20\% effect, on average.
Also our random errors are larger by $\sim$30\%, on average.
This overestimate of the bias could be due to the fact that the local
bias model slightly overestimates the linear bias, not taking into
account non-local contributions that are more significant at higher
orders, than at lower orders \citep[e.g. see][]{moresco2014,
  hoffmann2015}.  Nevertheless, our measurements are compatible at
68\% confidence level with the ones obtained from other probes. We
underline that the results obtained in this work are not only independent,
but also complementary to those obtained with other methods,
though all of them need to assume a model for dark matter clustering
statistics. Moreover, the strength of this analysis is
that the reduced 3PCF is independent of $\sigma_8$ (unlike the 2PCF),
and it is sensitive to the three-dimensional shape of cosmic structures
(unlike $S_3$).

The analysis on the $b_2$ parameter is less conclusive since it is less effectively constrained, 
with values in the range $-0.8<b_2<-0.2$, but relative errors of the order of $\sim40-60$ \%, on average.
The comparison with the measurements obtained from \cite{cappi2015}
shows a reasonable agreement for $M_{\rm B}\leq-20$, with the first
bin in particular presenting a significantly higher $b_2$. However, as also discussed in 
\cite{cappi2015}, the non-linear bias parameter is more difficult to constrain,
especially in the formalism used in that analysis, since it is extremely sensitive to errors on
$b_1$ and $S_3$ (see their eq. 20), and this might explain some of the scatter 
between the different measurements.
 A forthcoming 3PCF analysis on the final data release will allow us to further
investigate this issue, and to reduce the statistical errors. 


\section{Summary and conclusions}
\label{sec:concl}

In this paper, we investigated the dependence of higher-order
clustering on stellar mass and luminosity, providing measurements for
the first time at high redshift ($z\sim1$). We analysed galaxy samples
extracted from VIPERS PDR-1, in the redshift range $0.5<z<1.1$,
measuring both the connected and the reduced 3PCF in redshift space at
different scales.  The associated errors have been estimated from HOD
and SHMR mock catalogues, specifically constructed to reproduce VIPERS'
observational properties.
We provided measurements of the connected and reduced 3PCF as a
function of redshift, stellar mass, and luminosity for three
different scales, $r_{12}=2.5,5,10$ \Mpch, and $r_{13}=2\cdot r_{12}$, 
mapping from small to intermediate scales.

The main results of this work can be summarised as follows:

\begin{itemize}

\item We find a strong dependence of the reduced 3PCF on scales at all
  redshifts and for all stellar mass and luminosity bins, with an
  almost flat $Q(\theta)$ at smaller scales and a more prominent
  anisotropy at larger scales ($r_{12}=10$ \Mpch) independently of the redshift. 
  This trend can be interpreted as a signature of an increasing contribution of filamentary
  structures in the correlation function.

\item From the analysis of the connected 3CPF, $\zeta(\theta)$, we
  find that more massive and luminous galaxies present a stronger
  clustering, with a percentage difference of $\sim$20-40\%
    between the extreme bins, which is, however, not statistically
    relevant given the current uncertainties.  These results confirm
    the ones obtained at lower redshifts in SDSS, and extend them, for
    the first time, up to $z\sim1.1$. 

\item The reduced 3PCF, $Q(\theta)$, has a much smaller
  dependence on stellar mass and luminosity than $\zeta(\theta)$, with
  a percentage difference between the most massive and luminous bins
  of $\sim5$\%, in agreement with the results of previous studies.

\item We provide a first estimate of the linear bias parameter,
  $b_1$, exploiting the largest scales analysed in this work. The
  obtained constraints, in the range $1.5<b_1<2$, are compatible with
  previous measurements performed with independent approaches,
  considering both lower-order statistics or different estimators.

\end{itemize}

In a forthcoming paper we will take advantage of the higher statistics
provided by VIPERS final release to improve our constraints on the
3PCF, and to increase the accuracy on the linear and non-linear
  galaxy bias parameters.


\section*{Acknowledgments}

We acknowledge the crucial contribution of the ESO staff for the 
management of service observations. In particular, we are deeply 
grateful to M. Hilker for his constant help and support of this programme. 
Italian participation in VIPERS has been funded by INAF through PRIN 
2008 and 2010 programmes. LG and BRG acknowledge support of the European 
Research Council through the Darklight ERC Advanced Research Grant (\# 
291521). OLF acknowledges support of the European Research Council 
through the EARLY ERC Advanced Research Grant (\# 268107). AP, KM, and 
JK have been supported by the National Science Centre (grants 
UMO-2012/07/B/ST9/04425 and UMO-2013/09/D/ST9/04030), the Polish-Swiss 
Astro Project (co-financed by a grant from Switzerland, through the 
Swiss Contribution to the enlarged European Union). KM was supported by 
the Strategic Young Researcher Overseas Visits Program for Accelerating 
Brain Circulation No. R2405. RT acknowledges financial support from the European Research Council 
under the European Community's Seventh Framework Programme 
(FP7/2007-2013)/ERC grant agreement n. 202686. MM, EB, FM, and LM acknowledge the support from 
grants ASI-INAF I/023/12/0 and PRIN MIUR 2010-2011. LM also acknowledges 
financial support from PRIN INAF 2012. Research conducted within the scope of 
the HECOLS International Associated Laboratory, supported in part by the 
Polish NCN grant DEC-2013/08/M/ST9/00664.
We thank Hong Guo for kindly providing the data of their paper for comparison.

\bibliography{bib}


\appendix

\section{Covariance matrices} 
\label{appendix}
\begin{figure}[h!]
  \includegraphics[width=0.49\textwidth]{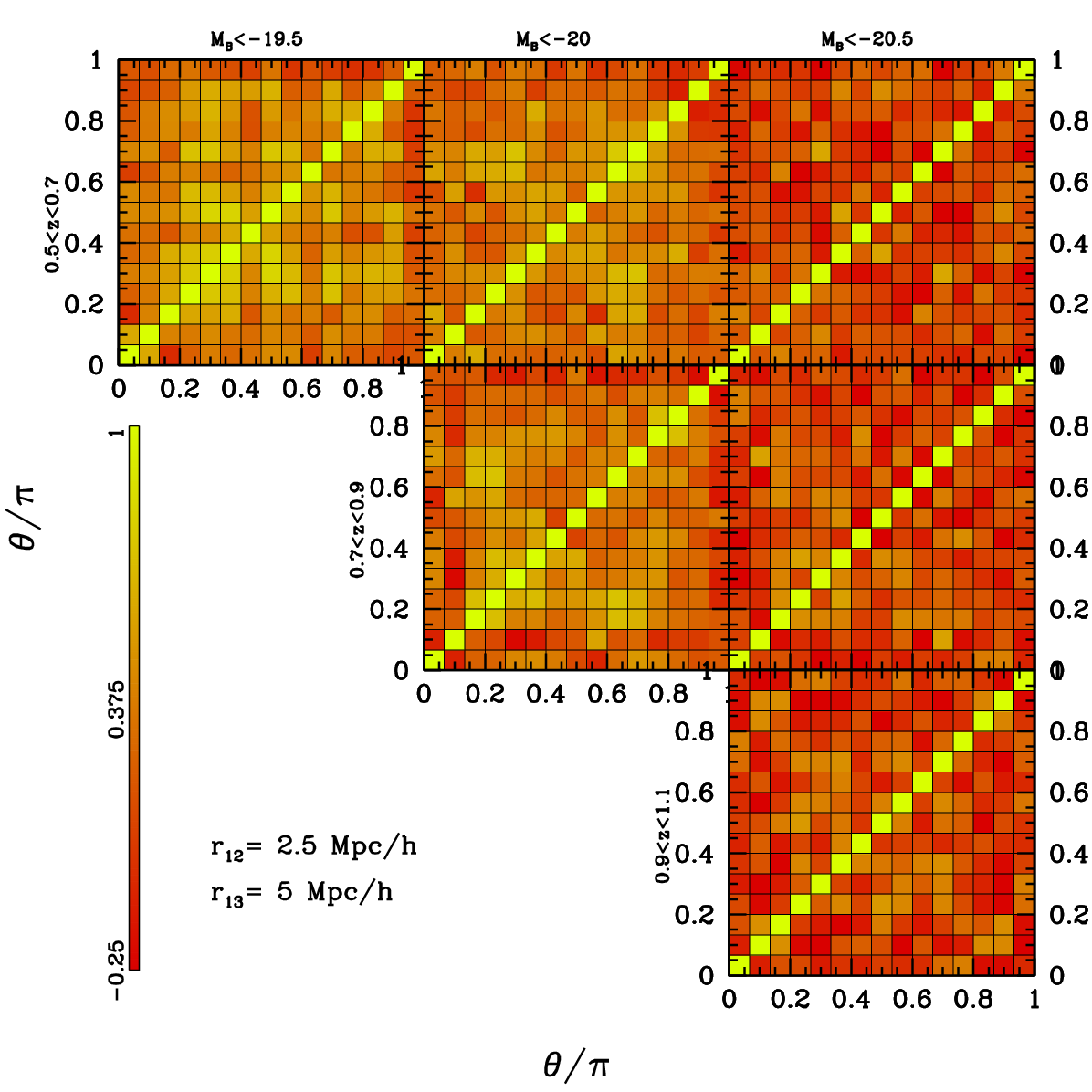}
  \caption{Normalised covariance matrices estimated at scales $r_{12}=2.5$
    \Mpch in the same luminosity and
  redshift bins as in Fig. \ref{fig:3PCF_2.5}.}
  \label{fig:cov1}
\end{figure}

\begin{figure}[h!]
  \includegraphics[width=0.49\textwidth]{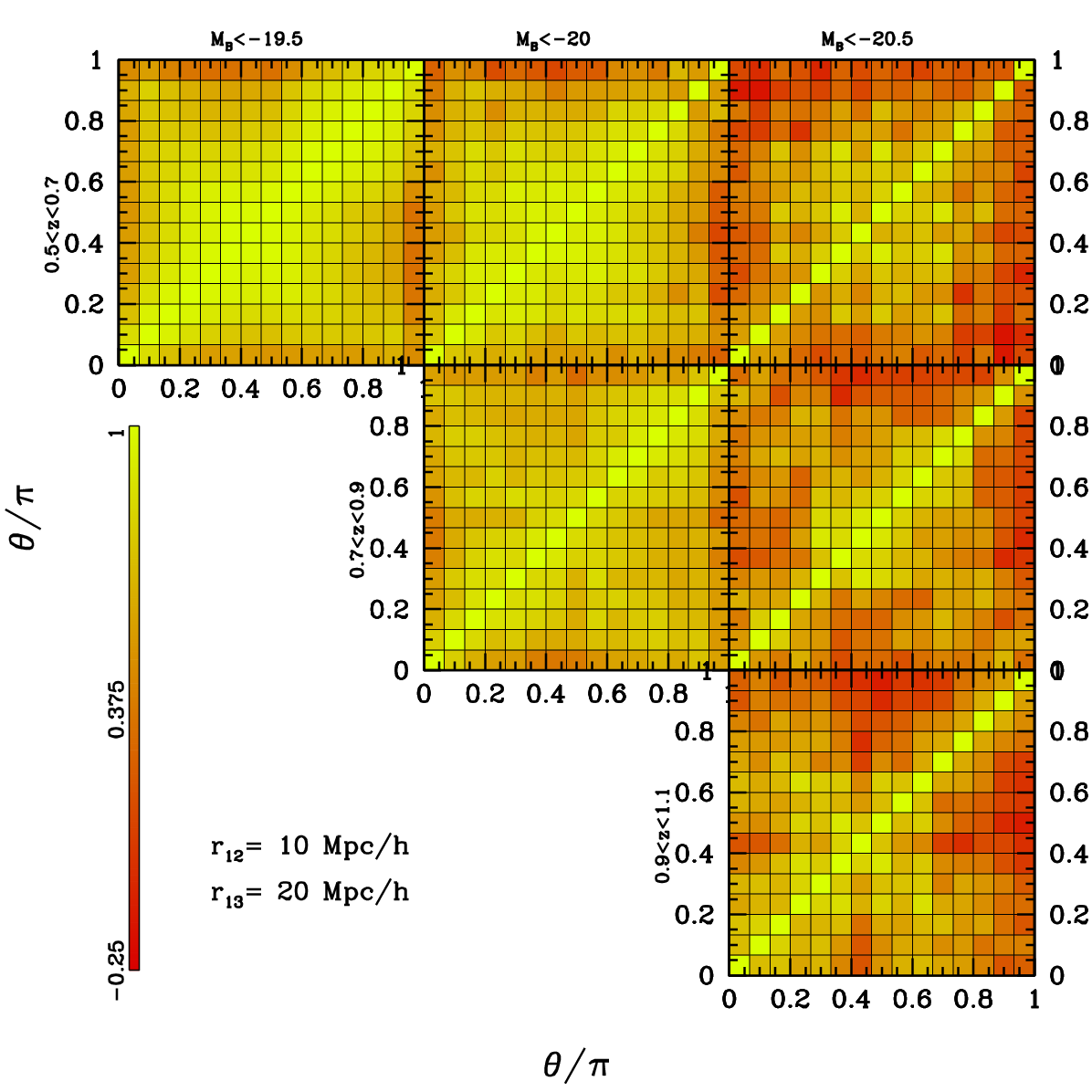}
  \caption{Same as Fig. \ref{fig:cov1}, but for scales $r_{12}=10$
    \Mpch.}
  \label{fig:cov2}
\end{figure}

In this analysis, the errors on both the connected and reduced 3PCF
are estimated from the diagonal elements of the covariance matrices,
calculated using the 26 mocks discussed in Sect. \ref{sub:errors}.
We have also estimated the full normalised covariance matrices, defined as:
\begin{equation}
\displaystyle C_{ij} =
\frac{1}{N}\sum_{k=1}^{N}\left(\frac{Q_{i}^{k}-\bar{Q}_{i}}{\sigma_{Q_{i}}}\right)
\left(\frac{Q_{j}^{k}-\bar{Q}_{j}}{\sigma_{Q_{j}}}\right) \, ,
\end{equation}
where $\sigma_{Q_{i}}$ is the value of the covariance of the i-th elements. 
We show for illustrative purposes in Figs. \ref{fig:cov1} and \ref{fig:cov2} the luminosity bins at two characteristic scales, $r_{12}=2.5$ and
$r_{12}=10$ \Mpch, respectively.

We find that at the smallest scales the covariance between bins is small,
decreasing with increasing luminosity threshold. At the highest scales
the correlation between bins is more significant, decreasing, also in this case,
with increasing luminosity threshold, and presents a similar
shape to the one obtained in other analyses \citep[e.g.][]{gaztanaga2005,hoffmann2015}.
Given the limited number of mocks available, in the analysis of this paper we have considered only 
the diagonal part of the covariance matrices, since off-diagonal elements may be less robustly estimated.
For the purpose of this paper this is not significant, and the main results and trends found 
are not affected by this assumption; we note, however, that their statistical significance might be
affected, especially at the largest scales.
In these cases, a statistical analysis of the data would require the use of
the full covariance to correctly estimate error-bars.
We plan to take it into account in a future analysis, in
which we plan to provide constraints on the galaxy bias.

\end{document}

%% file: authors.tex
\author{
M.~Moresco\inst{\ref{17},\ref{9}}
\and F.~Marulli\inst{\ref{17},\ref{9},\ref{18}}
\and L.~Moscardini\inst{\ref{17},\ref{9},\ref{18}}
\and E.~Branchini\inst{\ref{10},\ref{28},\ref{29}}
\and A.~Cappi\inst{\ref{9},\ref{30}}
\and I.~Davidzon\inst{\ref{4},\ref{9}}
\and B.~R.~Granett\inst{\ref{2}}
\and S.~de la Torre\inst{\ref{4}}
\and L.~Guzzo\inst{\ref{2},\ref{1}}
\and U.~Abbas\inst{\ref{5}}
\and C.~Adami\inst{\ref{4}}
\and S.~Arnouts\inst{\ref{4},\ref{6}}
\and J.~Bel\inst{\ref{31},\ref{32}}        
\and M.~Bolzonella\inst{\ref{9}}  
\and D.~Bottini\inst{\ref{3}}
\and C.~Carbone\inst{\ref{1}}
\and J.~Coupon\inst{\ref{33}}
\and O.~Cucciati\inst{\ref{9}}     
\and G.~De Lucia\inst{\ref{13}}
\and P.~Franzetti\inst{\ref{3}}
\and A.~Fritz\inst{\ref{3}}
\and M.~Fumana\inst{\ref{3}}
\and B.~Garilli\inst{\ref{3}}    
\and O.~Ilbert\inst{\ref{4}}
\and A.~Iovino\inst{\ref{2}}
\and J.~Krywult\inst{\ref{15}}
\and V.~Le Brun\inst{\ref{4}}
\and O.~Le F\`evre\inst{\ref{4}}
\and K.~Ma{\l}ek\inst{\ref{23}}
\and H.~J.~McCracken\inst{\ref{19}}
\and M.~Polletta\inst{\ref{3}}
\and A.~Pollo\inst{\ref{23},\ref{22}}
\and M.~Scodeggio\inst{\ref{3}} 
\and L.~A.~M.~Tasca\inst{\ref{4}}
\and R.~Tojeiro\inst{\ref{11}}
\and D.~Vergani\inst{\ref{25}}
\and A.~Zanichelli\inst{\ref{26}}
}
\offprints{M.Moresco \\ \email{michele.moresco@unibo.it}}
\institute{
Dipartimento di Fisica e Astronomia, Universit\`{a} di Bologna, Via Gobetti 93/2, I-40129, Bologna, Italy\label{17}
\and INAF - Osservatorio Astronomico di Bologna, via Ranzani 1, I-40127 Bologna, Italy \label{9}
\and INFN - Sezione di Bologna, viale Berti Pichat 6/2, I-40127 Bologna, Italy \label{18}
\and Dipartimento di Matematica e Fisica, Universit\`{a} degli Studi Roma Tre, via della Vasca Navale 84, 00146 Roma, Italy \label{10}
\and INFN - Sezione di Roma Tre, via della Vasca Navale 84, I-00146 Roma, Italy \label{28}
\and INAF - Osservatorio Astronomico di Roma, via Frascati 33, I-00040 Monte Porzio Catone (RM), Italy \label{29}
\and Laboratoire Lagrange, UMR7293, Universit\'e de Nice Sophia-Antipolis,  CNRS, Observatoire de la C\^ote d'Azur, 06300 Nice, France \label{30}
\and Aix Marseille Universit\'e, CNRS, LAM (Laboratoire d'Astrophysique de Marseille) UMR 7326, 13388, Marseille, France  \label{4}
\and INAF - Osservatorio Astronomico di Brera, Via Brera 28, 20122 Milano, via E. Bianchi 46, 23807 Merate, Italy \label{2}
\and Universit\`{a} degli Studi di Milano, via G. Celoria 16, 20133 Milano, Italy \label{1}
\and INAF - Osservatorio Astronomico di Torino, 10025 Pino Torinese, Italy \label{5}
\and Canada-France-Hawaii Telescope, 65--1238 Mamalahoa Highway, Kamuela, HI 96743, USA \label{6}
\and Aix Marseille Universit\'e, CNRS, CPT, UMR 7332, 13288 Marseille, France\label{31} 
\and Universit\'e de Toulon, CNRS, CPT, UMR 7332, 83957 La Garde, France\label{32}
\and INAF - Istituto di Astrofisica Spaziale e Fisica Cosmica Milano, via Bassini 15, 20133 Milano, Italy\label{3}
\and Department of Astronomy, University of Geneva ch. d'Ecogia 16, CH-1290 Versoix, Switzerland\label{33}
\and INAF - Osservatorio Astronomico di Trieste, via G. B. Tiepolo 11, 34143 Trieste, Italy \label{13}
\and Institute of Physics, Jan Kochanowski University, ul. Swietokrzyska 15, 25-406 Kielce, Poland \label{15}
\and National Centre for Nuclear Research, ul. Hoza 69, 00-681 Warszawa, Poland \label{23}
\and Institute d'Astrophysique de Paris, UMR7095 CNRS, Universit\'{e} Pierre et Marie Curie, 98 bis Boulevard Arago, 75014 Paris, France \label{19}
\and Astronomical Observatory of the Jagiellonian University, Orla 171, 30-001 Cracow, Poland \label{22}
\and Institute of Cosmology and Gravitation, Dennis Sciama Building, University of Portsmouth, Burnaby Road, Portsmouth, PO1 3FX, UK \label{11}
\and INAF - Istituto di Astrofisica Spaziale e Fisica Cosmica Bologna, via Gobetti 101, I-40129 Bologna, Italy \label{25}
\and INAF - Istituto di Radioastronomia, via Gobetti 101, I-40129 Bologna, Italy \label{26}
}